\documentclass[12pt]{article}
\SetSymbolFont{largesymbols}{normal}{OMX}{lcmex}{m}{n}
  
\catcode`@=11
\def\secteqno{\@addtoreset{equation}{section}%
\def\theequation{\thesection.\arabic{equation}}}
\catcode`@=12
\secteqno
\def\dd{\hbox{\,\Large$\triangleright$}}
\newcommand{\be}{\begin{equation}}
\newcommand{\ee}{\end{equation}}
\newcommand{\bea}{\begin{eqnarray}}
\newcommand{\eea}{\end{eqnarray}}
\newcommand{\bref}[1]{(\ref{#1})}
\newcommand{\nn}{\nonumber}

\newcommand{\slP}{/ {\hskip-0.27cm{P}}}
\newcommand{\slSigma}{/ {\hskip-0.27cm{\Sigma}}}

\newcommand{\sln}{/ {\hskip-0.27cm{n}}}

\def\dig#1{\setbox0=\hbox{$#1M$}
	\hskip.06\wd0 \vrule width.08\wd0 height.63\wd0 depth.01\wd0 
	\vrule width.37\wd0 height.63\wd0 depth-.55\wd0 \hskip-.4\wd0
	\vrule width.25\wd0 height.36\wd0 depth-.28\wd0 
	\vrule width.08\wd0 height.36\wd0 depth-.17\wd0 \hskip.14\wd0}
\def\digamma{{\mathpalette\dig{}}}

\begin{document}

\begin{flushright}
\parbox{4.2cm}
{2014,~November 9 \\
KEK-TH-1777 \hfill \\YITP-SB-14-42
\hfill \\
}
\end{flushright}

\vspace*{1.1cm}

\begin{center}
 {\Large\bf Ramond-Ramond gauge fields  }\\
 {~}\\
 {\Large\bf  in superspace with manifest T-duality  }
\end{center}
\vspace*{1.5cm}
\centerline{\large Machiko Hatsuda$^{\dagger^\natural }$\footnote{mhatsuda@post.kek.jp
}, Kiyoshi Kamimura$^{\ast }$\footnote{kamimura@ph.sci.toho-u.ac.jp}
and Warren Siegel$^\star$
\footnote{siegel@insti.physics.sunysb.edu
,
http://insti.physics.sunysb.edu/{\tt \~{}}siegel/plan.html}
}
\begin{center}
$^{\dagger}$\emph{Physics Division, Faculty of Medicine,
 Juntendo University, Chiba 270-1695, Japan}
\\
$^{\natural}$\emph{KEK Theory Center, High Energy Accelerator Research 
Organization,\\
Tsukuba, Ibaraki 305-0801, Japan} 
\\
$^{\ast}$\emph{Department of Physics, Toho University, Funabashi 274-8510, Japan
}\\
$^\star$\emph{C. N. Yang Institute for Theoretical Physics
State University of New York, Stony Brook, NY 11794-3840}
\vspace*{0.5cm}
\\

\end{center}

\vspace*{1cm}

\centerline{\bf Abstract}
\vspace*{0.5cm}
A superspace with manifest T-duality including Ramond-Ramond gauge fields
is presented. The superspace is defined by the double
nondegenerate super-Poincar\'{e}
algebras where  Ramond-Ramond charges are introduced by central extension.
This formalism allows a simple treatment that
all the supergravity multiplets are in a vielbein superfield and
all torsions with dimension 1 and less are trivial.
A Green-Schwarz superstring action is also presented
where  the Wess-Zumino term is given in a bilinear form of local currents.
Equations of motion are separated into left and right modes 
in a suitable gauge.
 \vfill 

\thispagestyle{empty}
\setcounter{page}{0}
\newpage

\section{Introduction}

T-duality is one of the most important  key ingredients
 of the gravity theory beyond general relativity
which may be the superstring theory.
The T-duality  was  realized manifestly by doubling 
coordinates firstly in \cite{Siegel:1993xq}.
T-duality covariant Lie bracket proposed in \cite{Siegel:1993xq}
 is turned out to be reduced into the Courant bracket
introduced in \cite{Hitchin:2004ut} to describe  generalized geometry.
Supersymmetric extension of T-duality symmetry 
with Ramond-Ramond (R-R) fields
 has been studied in 
\cite{Hassan:1999bv,Fukuma:1999jt,Hohm:2011zr,Jeon:2012kd}.
This subject is broaden in \cite{Hull:2004in,Grana:2005jc}
and its recent development is
in review articles for example 
\cite{Aldazabal:2013sca,Hohm:2013bwa,Berman:2014jba}.

A superspace approach of 
 type II supergravity theories 
  with manifest T-duality has been proposed 
in \cite{Hatsuda:2014qqa,Pola2014}.
Supergravity theories were constructed in superspace 
with almost all symmetries manifest  except T-duality symmetry in
\cite{Howe:1981gz,Howe:1983sra,Berkovits:2001ue}.
Our superspace involves two kinds of doubling in order to realize
manifest T-duality:
(1) { Nondegenerate algebra by introducing 
   nondegenerate pairs of generators such as $(D_\mu,~\Omega^\mu)$ \cite{Siegel:85, Siegel:1985xj,Siegel:1994xr} and $(S_{mn},~\Sigma^{mn})$ \cite{Hatsuda:2001xf,Bonanos:2009wy,Siegel:2011sy,Polacek:2013nla},
(2)  { Double coordinate space by using
 momenta and winding modes such as $(P_m,~P_{m'})$ \cite{Siegel:1993xq}.}
The doubling { (1)} is required from 
consistency of the stringy general coordinate invariance.
The Lorentz generator and its nondegenerate partner $(S_{mn},~\Sigma^{mn})$ are also 
basis  in order to make Lorentz connection to be
a part of the vielbein fields \cite{Polacek:2013nla}.
It may be useful for writing T-duality on the worldsheet. 
It is also possible to write 
the WZ term as bilinears of currents for arbitrary backgrounds.
This doubling is resolved by dimensional reduction constraints.
Double coordinate space {(2)} realizes a 
simpler and unified treatment of torsion constraints and vielbein. 
This doubling is resolved by imposing the section condition and
the strong constraint.

In this paper we include the R-R gauge fields
 to complete the construction of the superspace of the type II supergravity.
The R-R gauge fields are necessary as a type II superstring background, 
although a superstring does not carry R-R charges. 
The R-R gauge fields couple to  D$p$-branes, where $p$ is  even  for type IIA and
odd for type IIB theories.
Brane charges  are included by central extension of superalgebras  
\cite{Haag:1974qh,Siegel:1987ik}. 
Usually these brane algebras are realized on  worldvolumes.  
Their generalized  Lie bracket 
and gauge transformations of brane gauge fields  are obtained 
in \cite{Hatsuda:2012uk,Hatsuda:2012vm}.
In this paper the  R-R charges, denoted by $\Upsilon$, are introduced by  central extension of 
a superstring algebra.
Or equivalently $\Upsilon$ is introduced in an
affine algebra for the  double coordinate space by central extension. 
This central charge $\Upsilon$ is constrained to be zero on the superstring worldsheet,
but this is necessary to describe the  R-R gauge fields and its gauge symmetries.

Inclusion of the R-R fields distinguishes T-duality between IIA and IIB superstring theories in our formalism. 
Under a discrete T-duality transformation IIA theory and IIB theory are interchanged and
a D$p$-brane changes to a D$(p\pm 1)$-brane.
This is explained from the type II superstring algebra 
in double space as follows: 
There are two irreducible supertranslation algebras in a type II superstring theory
\bea
\{  {D}_{\mu}, {D}_{\nu}\}=2 {P}_m \gamma^m {}_{\mu\nu} &,&
~\{  {D}_{\mu'}, {D}_{\nu'}\}=2 {P}_{m'} \gamma^{m'} {}_{\mu'\nu'}\nn\\
{P}_m=p_{\rm m}+\partial_\sigma x^{\rm m}~&,&~
 {P}_{m'}=p_{\rm m}-\partial_\sigma x^{\rm m}~~~.\nn
\eea
Usual Lorentz vector index is denoted by $p_{\rm m}$.
Since in our approach two independent Lorentz generators are used,
 $S_{mn}$ for $P_m$ and $D_\mu$ and $S_{m'n'}$ for $P_{m'}$ and $D_{\mu'}$,
two chiralities of two chiral spinors $D_\mu$ and $D_{\mu'}$ are not fixed.
The type of theories, IIA or IIB, is determined after reducing two Lorentz symmetries into one Lorentz symmetry. 
A discrete T-duality transformation in  ${\bf n}$-th direction, 
which is an interchange between $p_{\bf n}$ and $\partial_\sigma x^{\bf n}$, is expressed in terms of 
an unit vector 
 $n^m=\delta^{m{\bf n}}$ and $n^{m'}=\delta^{m'{\bf n}}$ as
\bea
P_m\to P_m&,&{P}_{m'}\to  {P}_{m'}-  2n_{m'}~(n^{l'} {P}_{l'})~~~.\nn
\eea
It requires a change of chirality of $D_{\mu'}$ as
\bea
{D}_\mu  \to {D}_\mu
&,&
{D}_{\mu'}  \to i {D}_{\nu'}\sln^{\nu'\mu'}
~~,~~\sln=n_{n'}\gamma^{n'}~~~.\nn
\eea
The chirality of  $(D\sln)^{\mu'}$ is opposite to the one of 
$D_{\mu'}$, since $\sln$ anticommute with $\Gamma^{11}$.
This is a parity transformation which is part of O(d$-$1,1)$\times$O(d$-$1,1).
The  R-R charge $\Upsilon_{\mu\nu'}$
appears in a superalgebra between $D_\mu$ and $D_{\mu'}$.
The  R-R charge $\Upsilon_{\mu\nu'}$
is transformed under the discrete T-duality as
\bea
&
\{{D}_{{\mu}},{D}_{{\nu}'}\}
=2\Upsilon {}_{\mu\nu'}
~\to ~
\{{D}_{{\mu}},( {D}\sln)^{{\nu}'}\}
=2( \Upsilon ~\sln){}_{\mu}{}^{\nu'}  &\nn~~~.
\eea
The R-R charges $(\Upsilon \sln)_{\mu}{}^{\nu'}$ have 
different  form numbers from $\Upsilon_{\mu\nu'}$, 
when they are expanded by antisymmetric gamma matrix basis
after the dimensional reduction.
The form number is equal to $p$ for D$p$-brane.

In this paper we extend the manifestly T-duality formalism of type II superspace
to the one with the R-R gauge fields.
It requires the following procedures;
\begin{itemize}
  \item{Central extension of the affine Lie algebra}
   \item{Torsion constraints from $\kappa$-symmetric Virasoro constraints and Bianchi identities involving R-R charges}
   \item{Field identification and solving torsion constraints including   R-R fields} 
\end{itemize}
in addition to the previous procedures;
\begin{enumerate}
\item{Construct an affine Lie algebra and double the generators.}
\item{Make covariant derivatives with the vielbein field $E_{\underline{A}}{}^{\underline{M}}$.}
\item{Find torsion constraints from $\kappa$-symmetric Virasoro constraints and Bianchi identities.}
\item{Solve torsion constraints with constraints for dimensional reduction and the isotropy group
in addition to  the section conditions (the strong constraints).
\begin{enumerate}
\item{Section condition (strong constraints): 
The constraints reduces doubled coordinates to half, as in \bref{section}. 
This constraints gives rise to the stringy contribution in the ``new Lie derivative" 
\cite{Siegel:1993xq, Hatsuda:2014qqa}.
}
\item{Isotropy constraints:  The isotropy group constraints are usual gauge constarints for
a gauged coset model; such as Lorentz group part $S_{mn}=S_{m'n'}=0$ for Poincar$\acute{\rm e}$ group
\cite{Polacek:2013nla,Hatsuda:2014qqa}.}
\item{Dimensional reduction constraints: 
Dimensional reduction constraints are 
imposed on the unphysical symmetry currents
to remove auxiliary coordinates which are introduced to make nondegenerate group metric;
such as $\tilde{\Omega}^\mu=\tilde{\Omega}^{\mu'}=\tilde{\Sigma}^{mn}=\tilde{\Sigma}^{m'n'}=0$
\cite{Siegel:1994xr,Polacek:2013nla,Hatsuda:2014qqa,HKS14}.
}
\end{enumerate}
}
\end{enumerate}

The organization of the paper is the following:
In next section the R-R charges are introduced 
by a central extension of the affine Lie algebra.
In section 3 $\kappa$-symmetric Virasoro constraints involving R-R charges are obtained.
In section 4 torsion constraints are determined 
from the $\kappa$-symmetric Virasoro constraints and Bianchi identities.
 Components of the supervielbein field are 
identified with the supergravity contents including the R-R gauge fields.
Torsion constraints are solved.
In section 5 a Green-Schwarz superstring action in a manifestly T-duality formalism
is given. The Wess-Zumino term is given in a bilinear form of currents and 
equations of motion are
chirally separated in a suitable gauge.

\par \vskip 6mm

\section{Algebras}

In this paper we construct a superspace 
for the type II superstring background.
The superspace is defined by an affine nondegenerate 
double super-Poincar\'{e} algebras with 
R-R central charges.  
Notation follows our previous paper \cite{Hatsuda:2014qqa}.
After a central extension of a Lie algebra,
its affine extension is presented.

\subsection{Central extension of type II super-Poincar\'{e} algebra}

We begin by two independent sets of 
super-Poincar\'{e} algebras generated by $p_m,~s_{mn}$ with  
$_{m=0,1,\cdots,9}$ and $d_\mu$ with $_{\mu=1,\cdots,16}$ for left and
$p_{m'},~s_{m'n'}$ with   $_{m'=0,1,\cdots,9}$ 
and $d_{\mu'}$ with $_{\mu'=1,\cdots,16}$ for right.
A central extension of the double super-Poincar\'{e} algebras
is performed by introducing 
 R-R charges denoted by
``$\Upsilon_{\mu\nu'}$"
in an anticommutator between
 $d_\mu$ and $d_{\mu'}$. 
``$\Upsilon_{\mu\nu'}$"  carries mixed left and right indices.

Generators of two sets of super-Poincar\'{e} algebras including $\Upsilon_{\mu\nu'}$
are denoted by $G_I$.
Its affine extension requires nondegenerate metric.
The metric between $p_m$'s and $p_{m'}$'s are usual metric
$\eta_{mn}$ and $\eta_{m'n'}$.
In order to define a nondegenerate metric for $G_{I}$ except $p_m$ and $p_{m'}$ 
we introduce $G_{\tilde{I}}$:
\bea
{\renewcommand{\arraystretch}{1.6}
\begin{array}{|c||c||c||c||c|}
\hline
{\rm dim.}&0&\frac{1}{2}&1&1\\
G_{I}&s_{mn},~ s_{m'n'}&d_\mu,~d_{\mu'} &\Upsilon_{\mu\nu'}&p_{m},~p_{m'}
 \\\hline
 {\rm dim.}&2&\frac{3}{2}&1&\\
G_{\tilde{I}}&\sigma^{mn},~\sigma^{m'n'}&
\omega^\mu,~\omega^{\mu'}&\digamma^{\mu\nu'}&
\\\hline
\end{array}}
\label{bilinear}
\eea
The sum of  dimensions of a nondegenerate pair of
generators is 2.
Dimensions of $G_{I}$ are not
greater than 1, while dimensions of $G_{\tilde{I}}$ are not
less than 1.
The nondegenerate partner of 
the R-R central charge, $\Upsilon_{\mu\nu'}$, is  
 denoted by ``$\digamma^{\mu\nu'}$ (digamma)".

A double super-Poincar\'{e} algebra with central extension is 
  generated by $G_{I}$ 
 \bea
 &
{\renewcommand{\arraystretch}{1.6}
\begin{array}{lll}
{\rm dim}~0:&
\lbrack s_{mn},s_{lk}]=-i\eta_{[k|[m}s_{n]|l]}&
\lbrack s_{m'n'},s_{l'k'}]=i\eta_{[k'|[m'}s_{n']|l']}
\\
{\rm dim}~\frac{1}{2}:&
\lbrack s_{mn},d_{\mu}]=
-\frac{i}{2}(d\gamma_{mn})_\mu&
\lbrack s_{m'n'},d_{\mu'}]=
\frac{i}{2}(d\gamma_{m'n'})_{\mu'}
\\
{\rm dim}~1:&
\lbrack s_{mn},p_{l}]=-ip_{[m}\eta_{n]l}&
\lbrack s_{m'n'},p_{l'}]=ip_{[m'}\eta_{n']l'}
\\&\lbrack s_{mn},{\Upsilon}_{\mu\nu'}]=
-\frac{i}{2}{\Upsilon}_{\rho\nu'}(\gamma_{mn})^{\rho}{}_{\mu}
&
\lbrack s_{m'n'},{\Upsilon}_{\mu\nu'}]=
\frac{i}{2}{\Upsilon}_{\mu\rho'}(\gamma_{m'n'})^{\rho'}{}_{\nu'}\\
&\left\{d_{\mu},d_{\nu}\right\}
=2p_m \gamma^m{}_{\mu\nu}&
\left\{d_{\mu'},d_{\nu'}\right\}
=-2p_{m'} \gamma^{m'}{}_{\mu'\nu'}
\end{array}\nn
\label{RRalg0}}
&\\&~~~~~~
\{d_\mu,~d_{\nu'}\}~=~2\Upsilon_{\mu\nu'}&~~~.\label{Lie}
\eea
The gamma matrices satisfy 
\bea
(\gamma^m)_{\mu\nu}=(\gamma^m)_{\nu\mu}~,~
(\gamma^{(m|})^{\mu\rho}(\gamma^{|n)})_{\rho\nu}=2\eta^{mn}\delta^\mu_\nu
~,~(\gamma_{m})_{(\mu\nu}(\gamma^m)_{\rho)\lambda}=0\nn~~~.
\eea
 The rest of the commutators including generators of $G_{\tilde{I}}$ 
is given by
 \bea
& {\renewcommand{\arraystretch}{1.6}
\begin{array}{lll}
{\rm dim}~1:&
\lbrack s_{mn},\digamma^{\mu\nu'}]=
\frac{i}{2}(\gamma_{mn})^{\mu}{}_{\nu}
\digamma^{\nu\nu'}
&\lbrack s_{m'n'},\digamma^{\mu\nu'}]=
-\frac{i}{2}(\gamma_{m'n'})^{\nu'}{}_{\mu'}
\digamma^{\mu\mu'}  \\
{\rm dim}~\frac{3}{2}:&
\lbrack s_{mn},\omega^\mu]=
\frac{i}{2}(\gamma_{mn}\omega)^\mu & 
\lbrack s_{m'n'},\omega^{\mu'}]=
-\frac{i}{2}(\gamma_{m'n'}\omega)^{\mu'}\\&
\lbrack d_\mu,p_n]=2(\gamma_n\omega)_\mu & 
\lbrack d_{\mu'},p_{n'}]=-2(\gamma_{n'}\omega)_{\mu'}
\\&\lbrack d_\mu,\digamma^{\nu\nu'}]=
-2\delta^\nu_\mu\omega^{\nu'} 
&\lbrack d_{\mu'},\digamma^{\nu\nu'}]=
2\delta^{\nu'}_{\mu'}\omega^{\nu}  \\
{\rm dim}~2:&
\lbrack s_{mn},\sigma^{lk}]=-i\delta^{[k}_{[m}\sigma_{n]}{}^{l]} 
&
\lbrack s_{m'n'},\sigma^{l'k'}]=i\delta^{[k'}_{[m'}\sigma_{n']}{}^{l']} 
\\&
\left\{d_\mu,\omega^\nu\right\}=
-\frac{i}{4}\sigma^{mn}(\gamma_{mn})^\nu{}_\mu  &
\{d_{\mu'},\omega^{\nu'}\}=
\frac{i}{4}\sigma^{m'n'}(\gamma_{m'n'})^{n'}u{}_{\mu'}
\\&
\lbrack p_{m},p_{n}]=i\sigma_{mn} &
\lbrack p_{m'},p_{n'}]=-i\sigma_{m'n'}
\\
 
\end{array}}&\nn\\
&~~~~\lbrack \digamma^{\mu\nu'},
\Upsilon_{\nu\mu'}]=
\frac{i}{4}\left(
\delta^{\mu}_{\nu}\sigma^{m'n'}(\gamma_{m'n'})^{\nu'}{}_{\mu'}
+\delta^{\nu'}_{\mu'}\sigma^{mn}(\gamma_{mn})^{\mu}{}_{\nu}\right)&
\label{Lietilde}
\eea

\par
\vskip 6mm

\subsection{Central extension of affine type II super-Poincar\'{e} algebra}

An affine extension of 
 \bref{Lie} and \bref{Lietilde} is presented in this subsection. 
Generators of the affine Lie algebra 
are classified by the canonical dimensions as:
\bea
{\renewcommand{\arraystretch}{1.6}
\begin{array}{|c||c|c|c|c|c|c|}
\hline
{\rm dim.}&0&\frac{1}{2}&1&\frac{3}{2}&2\\
\mathring{\dd}{}_{\underline{M}}&S_{mn},~ S_{m'n'}&D_\mu,~D_{\mu'} &\Upsilon_{\mu\nu'},~P_{m},~P_{m'},~\digamma^{\mu\nu'}&\Omega^\mu,~\Omega^{\mu'}&\Sigma^{mn},~\Sigma^{m'n'}
\\\hline
\end{array}}
\eea
They satisfy
 \bea
\lbrack \mathring{\dd}{}_{\underline{M}}(1),\mathring{\dd}{}_{\underline{N}}(2)\}&=&
-if_{\underline{MN}}{}^{\underline{K}}\mathring{\dd}{}_{\underline{K}}\delta(2-1)
-i\eta_{\underline{MN}}\partial_\sigma\delta(2-1)~~~,\label{tratraf}
\eea
where the structure constant $f_{\underline{MN}}{}^{\underline{K}}$ is given by
\bref{RRalg0} and \bref{Lietilde}.
The nondegenerate metric $\eta_{\underline{MN}}$ is given as:
\bea
&\eta_{\underline{MN}}
={\renewcommand{\arraystretch}{1.4}
\begin{array}{c}S\\D\\{\Upsilon}
\\P\\{\digamma}\\\Omega\\\Sigma\end{array}
\left(\begin{array}{ccccccc}
& & &&& &{\bf 1}_S\\
 & &&& &{\bf 1}_D& \\
 &&&&{\bf 1}_\Upsilon&&\\
 && &{\bf 1}_P& && \\
 &&{\bf 1}_\Upsilon{}&&&&
 \\
 &-{\bf 1}_D{}& &&& & \\
{\bf 1}_S{}& & &&& & 
\end{array}\right)}~~\label{etaMNRR}
~,~
{\renewcommand{\arraystretch}{1.4}
\begin{array}{l}
{\bf 1}_S=
\left(\begin{array}{cc}\delta_{[m}^l\delta_{n]}^{k}&\\
& -\delta_{[m'}^{l'}\delta_{n']}^{k'}\end{array}\right)\\
{\bf 1}_D=\left(\begin{array}{cc}  \delta_\mu^\nu &\\
&- \delta_{\mu'}^{\nu'}\end{array}\right)\\
{\bf 1}_\Upsilon=\delta^\nu_{\mu}\delta^{\nu'}_{\mu'}\\
{\bf 1}_P=\left(\begin{array}{cc}\eta_{mn}&\\
&-\eta_{m'n'} \end{array}\right)
\end{array}}&\nn
\eea

The affine nondegenerate super-Poincar\'{e} algebra in components
 is given by
 \bea
{\renewcommand{\arraystretch}{1.6}
\begin{array}{llcl}
{\rm dim }~0:&\lbrack S_{mn}(1),S_{lk}(2)]&=&-i\eta_{[k|[m}S_{n]|l]}\delta(2-1)\\
{\rm dim }~\frac{1}{2}:&\lbrack S_{mn}(1),D_{\mu}(2)]&=&-\frac{i}{2}(D\gamma_{mn})_\mu\delta(2-1)\\
{\rm dim }~1:&\lbrack S_{mn}(1),P_{l}(2)]&=&-iP_{[m}\eta_{n]l}\delta(2-1)\\
&\left\{D_\mu(1),D_\nu(2)\right\}&=&2P_m \gamma^m{}_{\mu\nu}\delta(2-1)\\
{\rm dim }~\frac{3}{2}:&\lbrack S_{mn}(1),\Omega^\mu(2)]&=&
\frac{i}{2}(\gamma_{mn}\Omega)^\mu \delta(2-1)\\
&\lbrack D_\mu(1),P_n(2)]&=&2(\gamma_n\Omega)_\mu \delta(2-1)\\
{\rm dim }~2:&\lbrack S_{mn}(1),\Sigma^{lk}(2)]&=&-i\delta^{[k}_{[m}\Sigma_{n]}{}^{l]}\delta(2-1)
+i\delta_{[m}^l\delta_{n]}^k\partial_\sigma\delta(2-1)
\\
&\left\{D_\mu(1),\Omega^\nu(2)\right\}&=&
-\frac{i}{4}\Sigma^{mn}(\gamma_{mn})^\nu{}_\mu \delta(2-1)
+i\delta_\mu^\nu \partial_\sigma\delta(2-1)\\
&\lbrack P_{m}(1),P_{n}(2)]&=&i\Sigma_{mn}\delta(2-1)
+i\eta_{mn}\partial_\sigma\delta(2-1)
\end{array}}\label{SDPomesig}
\eea
Commutators with  dimension greater than 2 are zero.
The right currents $(S',D',P',\Omega',\Sigma')$
satisfy the same algebra with opposite sign in the right hand sides.

Algebras involving the R-R charges $\Upsilon_{\mu\mu'}$
and $\digamma^{\mu\mu'}$ are the followings:
 \bea
{\renewcommand{\arraystretch}{1.6}
\begin{array}{llcl}
{\rm dim }~1:&\lbrack S_{mn}(1),\digamma^{\mu\nu'}(2)]&=&
\frac{i}{2}(\gamma_{mn})^{\mu}{}_{\nu}
\digamma^{\nu\nu'}\delta(2-1)\\
&\lbrack S_{m'n'}(1),\digamma^{\mu\nu'}(2)]&=&
-\frac{i}{2}(\gamma_{m'n'})^{\nu'}{}_{\mu'}
\digamma^{\mu\mu'}\delta(2-1)\\
&\lbrack S_{mn}(1),{\Upsilon}_{\mu\nu'}(2)]&=&
-\frac{i}{2}{\Upsilon}_{\nu\nu'}(\gamma_{mn})^{\nu}{}_{\mu}
\delta(2-1)\\
&\lbrack S_{m'n'}(1),{\Upsilon}_{\mu\nu'}(2)]&=&
\frac{i}{2}{\Upsilon}_{\mu\mu'}(\gamma_{m'n'})^{\mu'}{}_{\nu'}
\delta(2-1)\\
&\left\{D_\mu(1),D_{\nu'}(2)\right\}&=&
2\Upsilon_{\mu\nu'}\delta(2-1)\\%
{\rm dim }~\frac{3}{2}:
&\lbrack D_\mu(1),\digamma^{\nu\nu'}(2)]&=&
-2\delta^\nu_\mu\Omega^{\nu'} \delta(2-1)\\
&\lbrack D_{\mu'}(1),\digamma^{\nu\nu'}(2)]&=&
2\delta^{\nu'}_{\mu'}\Omega^{\nu} \delta(2-1)\\
{\rm dim }~2:&\lbrack \digamma^{\mu\nu'}(1),
\Upsilon_{\nu\mu'}(2)]&=&
\frac{i}{4}\left(
\delta^{\mu}_{\nu}\Sigma^{m'n'}(\gamma_{m'n'})^{\nu'}{}_{\mu'}
+\delta^{\nu'}_{\mu'}\Sigma^{mn}(\gamma_{mn})^{\mu}{}_{\nu}\right)
\delta(2-1)\\
&&&+i\delta_\nu^\mu\delta_{\mu'}^{\nu'}\partial_\sigma(2-1)
\end{array}}\label{RRalg}
\eea
Commutators with  dimension greater than 2 are zero.

\par
\vskip 6mm

\section{
{$\kappa$}
-symmetric Virasoro constraints}

As the Virasoro constraint restricts string backgrounds,
the $\kappa$-symmetry constraint together with the Virasoro constraint restrict 
superstring backgrounds \cite{Witten:1985nt}.
The $\kappa$-symmetry at the first quantized level  guarantees the number of 
degrees of freedom of fermionic coordinates,
while the $\kappa$-symmetric Virasoro constraint 
guarantees the physical degrees of freedom of 
 the background fields. 
 It was also shown that 
 the $\kappa$-symmetric Virasoro algebra restricts
torsions in a superspace \cite{Shapiro:1986xp}.
This restriction was presented for the manifestly T-dual type II superspace \cite{Hatsuda:2014qqa}.
 In this section consistency of the $\kappa$-symmetry is examined from the fermionic constraints
of the Green-Schwarz superstring.
Then $\kappa$-symmetric Virasoro constraints are obtained.

\subsection{$\kappa$-symmetry}

In the Green-Schwarz formalism fermionic constraints, 
$D_{\underline{\mu}}=(D_\mu,~D_{\mu'})=0$, 
are mixed constraints of first class 
and second class, where the first class constraints generate the $\kappa$-symmetry. 
In order to preserve this structure an anticommutator between 
two fermionic constraints must have zero determinant,
\bea
\{D_{\underline{\mu}},D_{\underline{\nu}}\}&=&\Xi_{\underline{\mu\nu}}~~,~~
\Xi_{\underline{\mu\nu}}~=~
2\left(
\begin{array}{cc}
\slP_{\mu\nu}&\Upsilon_{\mu\nu'}\\
\Upsilon_{\nu\mu'}&-\slP_{\mu'\nu'}
\end{array}
\right)~~,~~
\det \Xi_{\underline{\mu\nu}}=0~~~
\eea
with $\slP=P_m\gamma^m$.
$\Xi_{\underline{\mu\nu}}$ must  have rank 16 and it has 16 independent zero eigen vectors. 
Zero eigen matrix of $\Xi_{\underline{\mu\nu}}$ is chosen with a constant  $\alpha$ as 
\bea
\bar{\Xi}^{\underline{\mu\nu}}~=~
\left(
\begin{array}{cc}
\slP^{\mu\nu}&\alpha\digamma^{\mu\nu'}\\
\alpha\digamma^{\nu\mu'}&\slP^{\mu'\nu'}
\end{array}
\right)~,
\eea
then it satisfies 
\bea
&\Xi_{\underline{\mu\nu}}\bar{\Xi}^{\underline{\nu\lambda}}~\approx~0
~~\Leftrightarrow ~~
{\renewcommand{\arraystretch}{1.8}
\left\{\begin{array}{l}
P_mP^m+\frac{\alpha}{16}{\rm tr}\Upsilon \digamma=0\\
P_{m'}P^{m'}-\frac{\alpha}{16}{\rm tr}\Upsilon \digamma=0\\
\Upsilon_{\mu\nu'}\digamma^{\nu\nu'}|_{\rm trls}=
\Upsilon_{\mu\mu'}\digamma^{\mu\nu'}|_{\rm trls}=0\\
\Upsilon_{\mu\mu'} \slP^{\mu'\nu'}+\alpha\slP_{\mu\nu}\digamma^{\nu\nu'}=0\\
\slP^{\mu\nu}\Upsilon_{\nu\nu'}-\alpha\digamma^{\mu\mu'}\slP_{\mu'\nu'}=0
\end{array}\right.}&\label{kappaconst}
\eea
The first and second lines of \bref{kappaconst} are Klein-Gordon operators and 
the last two lines are Dirac-K\"{a}hler operators.
The term ${\rm tr}\Upsilon \digamma$ becomes BPS mass for D-branes
where $\Upsilon$ and $\digamma$ take D-brane volume forms \cite{Hatsuda:2012uk,Hatsuda:2012vm}.
The $\kappa$-symmetry constraints are obtained as 
\bea
{\cal B}^{\underline{\mu}}_0&=&D_{\underline{\nu}}\bar{\Xi}^{\underline{\nu\mu}}~=
{\renewcommand{\arraystretch}{1.8}
\left\{
\begin{array}{l}
{\cal B}^\mu_0=D_{\nu}\slP^{\nu\mu}+\alpha D_{\nu'}\digamma^{\mu\nu'}=0\\
{\cal B}^{\mu'}_0=D_{\nu'}\slP^{\nu'\mu'}+\alpha D_{\nu}\digamma^{\nu\mu'}=0\\
\end{array}
\right.}\label{Bkappa}
\eea
so that they are first class 
\bea
\{{\cal B}_0^{\underline{\mu}},{\cal B}_0^{\underline{\nu}}\}=2
\bar{\Xi}^{\underline{\mu\rho}}~{\Xi}_{\underline{\rho\lambda}}~\bar{\Xi}^{\underline{\lambda\nu}}
+\left(D,~DD ~{\rm constraints}\right)\approx 0~~~.
\eea

In this paper we focus on background solutions which couple to superstrings
not D-branes, so we impose an additional constraint
\bea
\Upsilon_{\mu\nu'}=0\label{Upsilonconstraint}~~~.
\eea
Consistency with \bref{kappaconst} and \bref{Upsilonconstraint}, 
$\{\Upsilon,\Xi\bar{\Xi}\}\approx 0$,
leads to $\alpha=0$. 
This gives massless conditions: 
\bea
&\Xi_{\underline{\mu\nu}}\bar{\Xi}^{\underline{\nu\lambda}}~\approx~0
~~\Leftrightarrow ~~
{\renewcommand{\arraystretch}{1.8}
\left\{\begin{array}{l}
P_mP^m=0\\
P_{m'}P^{m'}=0
\end{array}\right.}&\label{kappaconstst}
\eea
Although it reduces to the same algebra with the $\kappa$-symmetric Virasoro algebra
without central extension, $\digamma$ is necessary to introduce the R-R gauge field.
\par\vskip 6mm

\subsection{$\kappa$-symmetric Virasoro constraints}

The  $\sigma$ -diffeomorphism constraint is given by
\bea
{\cal H}_\sigma&=&\frac{1}{2}
\mathring{\dd}{}_{\underline{M}}\eta^{\underline{M}\underline{N}}\mathring{\dd}{}_{\underline{N}}\label{Virsig}
\\
&=&
\frac{1}{2}P_mP^m-\frac{1}{2}P_{m'}P^{m'}
+\Omega^{\mu}D_\mu-\Omega^{\mu'}D_{\mu'}+
\frac{1}{2}\Sigma^{mn}S_{mn}-\frac{1}{2}\Sigma^{m'n'}
S_{m'n'}+\digamma^{\mu\nu'}\Upsilon_{\mu\nu'}~~~\nn
\eea
satisfying
\bea
[{\cal H}_\sigma(\sigma),{\cal H}_\sigma(\sigma')]&=&-i\left(
2{\cal H}_\sigma(\sigma)\partial_\sigma\delta(\sigma-\sigma')+
\partial_\sigma{\cal H}_\sigma \delta(\sigma-\sigma')
\right)~~~.\label{sigsig}
\eea
The commutator with ${\cal H}_\sigma$ gives $\sigma$ derivatives of operators,
for example
$\left[{\cal O}(\sigma),i\int {\cal H}_\sigma\right]
=\partial_\sigma {\cal O}(\sigma)$. 
The $\tau$-diffeomorphism constraint is given by
\bea
{\cal H}_\tau&=&\frac{1}{2}
\mathring{\dd}{}_{\underline{M}}\hat{\eta}^{\underline{M}\underline{N}}
\mathring{\dd}{}_{\underline{N}}
\\
&=&
\frac{1}{2}P_mP^m+\frac{1}{2}P_{m'}P^{m'}
+\Omega^\mu D_{\mu}+\Omega^{\mu'} D_{\mu'}+
\frac{1}{2}\Sigma^{mn}S_{mn}+\frac{1}{2}\Sigma^{m'n'}S_{m'n'}~~~\nn
\eea
where a metric $\hat{\eta}_{\underline{MN}}$ is defined as:
\bea
&\hat{\eta}_{\underline{MN}}
={\renewcommand{\arraystretch}{1.4}
\begin{array}{c}S\\D\\{\Upsilon}
\\P\\{\digamma}\\\Omega\\\Sigma\end{array}
\left(\begin{array}{ccccccc}
& & &&& &\hat{\bf 1}_S\\
 & &&& &\hat{\bf 1}_D& \\
 &&&&{\bf 0}&&\\
 && &\hat{\bf 1}_P& && \\
 &&{\bf 0}&&&&
 \\
 &-\hat{\bf 1}_D{}& &&& & \\
\hat{\bf 1}_S{}& & &&& & 
\end{array}\right)}~~&\label{etahat}
~,~
{\renewcommand{\arraystretch}{1.4}
\begin{array}{l}
\hat{\bf 1}_S=
\left(\begin{array}{cc}\delta_{[m}^l\delta_{n]}^{k}&\\
& \delta_{[m'}^{l'}\delta_{n']}^{k'}\end{array}\right) \\
\hat{\bf 1}_D=\left(\begin{array}{cc}  \delta_\mu^\nu &\\
& \delta_{\mu'}^{\nu'}\end{array}\right)  \\
\hat{\bf 1}_P=\left(\begin{array}{cc}\eta_{mn}&\\
&\eta_{m'n'} \end{array}\right)
\end{array}}
\eea
 ${\cal H}_\sigma={\cal H}_\tau=0$ are  
essentially massless constraints in two spaces 
given in \bref{kappaconstst}
if we use first and second class constraints, $D_\mu=D_{\mu'}=0$ and 
$S_{mn}=S_{m'n'}=\Upsilon_{\mu\nu'}=0$.
They satisfy the following Virasoro algebra
\bea
 \lbrack{\cal H}_\sigma(1),{\cal H}_\tau(2)]&=&i
({\cal H}_\tau(1)+{\cal H}_\tau(2))\partial_\sigma \delta(2-1)\nn\\
 \lbrack{\cal H}_\tau(1),{\cal H}_\tau(2)]&=&i
 \left\{({\cal H}_\sigma-{\rm tr}\digamma\Upsilon)(1)+({\cal H}_\sigma-{\rm tr}\digamma\Upsilon)(2)\right\}
 \partial_\sigma \delta(2-1)~~~\nn\\
 \lbrack{\cal H}_\sigma(1),\Upsilon_{\mu\nu'}(2)]&=&i\Upsilon_{\mu\nu'}(1)\partial_\sigma\delta(2-1)
+ \frac{i}{4}(\delta^\rho_\mu\slSigma^{\rho'}{}_{\nu'}
+\delta^{\rho'}_{\nu'}\slSigma^\rho{}_{\mu} )\Upsilon_{\rho\rho'}\delta(2-1)
\nn
 \\\lbrack{\cal H}_\tau(1),\Upsilon_{\mu\nu'}(2)]&=&
 0
\eea
with $\slSigma=\Sigma_{mn}\gamma^{mn}$. 
Other constraints except ${\cal H}_\sigma$ 
do not include $\digamma^{\mu\nu'}$ so they commute with $\Upsilon$, such as
$\Upsilon_{\mu\nu'}=0$ and $S_{mn}=S_{m'n'}=0$. 

A consistent set of $\kappa$-symmetric Virasoro constraints are
\bea
\left(
{\cal H}_\tau,~{\cal H}_\sigma,~{\cal B}^{\mu},~{\cal B}^{\mu'}
,~S_{mn},~S_{m'n'},~\Upsilon_{\mu\nu'},~{\cal C}_{\mu\nu},~
{\cal C}_{\mu'\nu'}
,~{\cal D}_m,~{\cal D}_{m'}\right)=0
~~~.\label{abcd}
\eea
The $\kappa$-symmetric Virasoro algebras for flat left and flat right 
with $\Upsilon$ and $\digamma$ are given as
\bea
{\cal A}&=&\frac{1}{2}P_mP^m+\Omega^\mu D_\mu
+\frac{1}{2}\Sigma^{mn}S_{mn}+\frac{1}{2}\digamma^{\mu\nu'}\Upsilon_{\mu\nu'}
\nn\\
{\cal B}^\mu&=&
D_\nu\slP^{\nu\mu}-iS_{mn}(\gamma^{mn}\Omega)^\mu\nn\\
{\cal A}'&=&\frac{1}{2}P_{m'}P^{m'}+\Omega^{\mu'} D_{\mu'}
+\frac{1}{2}\Sigma^{m'n'}S_{m'n'}-\frac{1}{2}\digamma^{\mu\nu'}\Upsilon_{\mu\nu'}\nn\\
{\cal B}^{\mu'}&=&D_{\nu'}\slP^{\nu'\mu'}-iS_{m'n'}(\gamma^{m'n'}\Omega)^{\mu'}
\nn
\eea
 ${\cal CD}$ constraints are constructed by bilinears of $D_\mu$ and $D_{\mu'}$.
${\cal ABCD}$ constraints  make a closed algebra as same as the one without central extension
up to the $\Upsilon=0$ constraints given in \cite{Hatsuda:2014qqa}.
${\cal ABCD}$ constraints are extended to curved backgrounds
as shown in \cite{Hatsuda:2014qqa}.
\par\vskip 6mm

\section{Torsion and vielbein}

Now we  examine  superstring backgrounds by introducing 
the vielbein field.
Simplification occurs by the central extension of $\Upsilon_{\mu\nu'}$
among  several results following our previous paper \cite{Hatsuda:2014qqa}.
The vielbein field includes the R-R gauge fields as well as 
R-R field strengths.
Gauge transformation rules and solutions of torsion constraints are presented.
We identify type II supergravity fields to  
vielbein field components $E_{\underline{A}}{}^{\underline{M}}$. 

\subsection{Torsion constraints}

Covariant derivatives in curved backgrounds are written in terms of vielbein fields as
\bea
\dd_{\underline{A}}&=&E_{\underline{A}}{}^{\underline{M}}(Z^{\underline{N}})\mathring{\dd}{}_{\underline{M}}\nn\\
  \lbrack\dd_{\underline{A}},\dd_{\underline{B}}\}&=&-iT_{\underline{AB}}{}^{\underline{C}}\dd_{\underline{C}}\delta(2-1)
  -i\eta_{\underline{AB}}\partial_\sigma\delta(2-1)~~~.\label{affinealge}
\eea
The manifestly T-dual formulation allows to impose
an orthonormal condition on the vielbein field.
It results the second term of the right hand side of \bref{affinealge} as
\bea
E_{\underline{A}}{}^{\underline{M}}\eta_{\underline{MN}}E_{\underline{B}}{}^{\underline{N}}=\eta_{\underline{AB}}~~,~~
\eta_{\underline{MN}}=\eta_{\underline{AB}}~~~.
\eea
Torsions and  Bianchi identities with lower indices are totally graded antisymmetric
\bea
T_{\underline{ABC}}&=&T_{\underline{AB}}{}^{\underline{D}}\eta_{\underline{DC}}
=\frac{1}{3!}T_{[\underline{ABC})}=
\frac{1}{2}(D_{[\underline{A}}E_{\underline{B}}{}^{\underline{M}})E_{\underline{C})\underline{M}}+E_{\underline{A}}{}^{\underline{M}}E_{\underline{B}}{}^{\underline{N}}
E_{\underline{C}}{}^{\underline{L}}f_{\underline{MNL}}\nn\\
{\cal I}_{\underline{ABCD}}&=&\frac{1}{4!}{\cal I}_{[\underline{ABCD})}
=D_{[\underline{A}}T_{\underline{BCD})}
+\frac{3}{4}T_{[\underline{AB}}{}^{\underline{E}}T_{\underline{CD})\underline{E}}=0\label{Bian}
\eea 
with $D_{\underline{A}}=E_{\underline{A}}{}^{\underline{M}}D_{\underline{M}}$
and 
$f_{\underline{MNL}}=f_{\underline{MN}}{}^{\underline{K}}\eta_{\underline{KL}}$
 is the structure constant in \bref{tratraf}.
Graded antisymmetrization is denoted as ${\cal O}_{[AB)}={\cal O}_{AB}-(-)^{AB}{\cal O}_{BA}$.

The Bianchi identity and the consistency of the $\kappa$-symmetry determine 
a set of torsion constraints.
Indices are abbreviated as
$V_a=V_{P_a}$, $V_{\underline{a}}=(V_{P_a},V_{P_{a'}})$, 
$V_{\alpha}=V_{D_\alpha}$, $V_{\underline{\alpha}}=(V_{D_\alpha},V_{D_{\alpha'}})$.
We impose torsion constraints for Lorentz indices as
\bea
T_{\underline{SAB}}=f_{\underline{SAB}}~~~.
\eea
Then torsions with two $P_a$'s indices, $T_{ab}{}^{\underline{M}}$, 
are absorbed into $E_{\Sigma}{}^{\underline{M}}$ as
\bea
&&[\dd_{a},\dd_b]=T_{ab}{}^{\underline{A}}\dd_{\underline{A}}
~\to~
f_{ab}{}^{\Sigma}\dd_\Sigma= f_{abS}
E_\Sigma{}^{\underline{M}}Z_{\underline{M}}\label{TUpzero}~~~.
\eea
For superspaces with fermionic constraints, most of torsions with dimension lower than 1 are   structure constants or zero.
A supercurvature $T_{\alpha \beta\Sigma}$ 
is absorbed into vielbein $E_{c\Sigma}$ as
$f_{\alpha\beta}{}^c E_{c\Sigma}$.
The supercurvature $R_{\alpha\beta'}{}^{cd}=T_{\alpha\beta\Sigma}$,
 which was nonzero for AdS space,
is absorbed into vielbein after introducing $\Upsilon_{\mu\nu'}$ as 
\bea
\{\dd_\alpha,\dd_{\beta'}\}=T_{\alpha \beta'\Sigma}{}{S}~\to~
f_{\alpha\beta'}{}^{\Upsilon}E_{\Upsilon\Sigma}{} S~~~.
\eea
A dimension 1 torsion $T_{\alpha' a}{}^\gamma$, 
which was nonzero without $\Upsilon_{\mu\nu'}$,
becomes zero from the Bianchi identity ${\cal I}_{\alpha\alpha' ab}=0$ in \bref{Bian}
with help of $T_{ab\Upsilon}=0$ in \bref{TUpzero}.

Trivial and non-trivial torsions are listed as follows:
\bea
{\renewcommand{\arraystretch}{1.4}
\begin{array}{c|c|c}
{\rm dim}&{\rm zero}&{\rm nonzero}\\\hline
-1/2&T_{\alpha\beta{\gamma}},T_{\alpha\beta\gamma'}
&{\rm NONE}\\\hline
0&
\begin{array}{c}
T_{\alpha\beta c}-f_{\alpha\beta c},~T_{\alpha\beta' \digamma}-f_{\alpha\beta' \digamma}\\
T_{abS}-f_{abS},~T_{\alpha}{}^{\beta}{}_{S}-f_{\alpha}{}^{\beta}{}_{S}\\
T_{\Upsilon\digamma \underline{S}}-f_{\Upsilon\digamma \underline{S}},~
T_{\alpha\underline{\beta} \Upsilon}
,T_{\alpha}{}^{\beta'}{}_{\underline{S}} \\
T_{\alpha\beta c'},T_{\alpha\beta' c},
T_{\alpha\beta \digamma},
T_{ab'S},\\
\end{array}&{\rm NONE}\\\hline
1/2&
\begin{array}{c}
T_{\alpha\underline{\beta}}{}^{\underline{\gamma}},
T_{\alpha b \underline{c}},T_{\alpha b'c'}
,
T_{\alpha \underline{b} \Upsilon},T_{\alpha \underline{b} \digamma}\\
,
T_{\alpha \Upsilon\Upsilon},T_{\alpha \Upsilon\digamma},
T_{\alpha \digamma\digamma}\\
\end{array}
&{\rm NONE}\\\hline
1&
\begin{array}{c}
T_{\alpha \underline{b}}{}^{\underline{\gamma}},
T_{\alpha \Upsilon}{}^{\underline{\gamma}},
T_{\alpha \digamma}{}^{\underline{\gamma}},
T_{\alpha\underline{\beta}\underline{\Sigma}},
\\
T_{ab\underline{c}},
T_{a\underline{b}\Upsilon},
T_{a\underline{b}\digamma},
T_{a\Upsilon\Upsilon},T_{a\Upsilon\digamma},
T_{a\digamma\digamma}
\\T_{\Upsilon\Upsilon\Upsilon},T_{\Upsilon\Upsilon\digamma},
T_{\Upsilon\digamma\digamma},T_{\digamma\digamma\digamma}
\end{array}&{\rm NONE}\\\hline
3/2&
T_{\alpha}{}^{\underline{\beta}\underline{\gamma}},
T_{ab}{}^{\underline{\gamma}},
T_{\alpha \underline{b} \underline{\Sigma}},T_{\alpha \digamma\Sigma},T_{a \digamma}{}^{\beta},
T_{a\Upsilon}{}^{\beta'},T_{\Upsilon\digamma}{}^{\underline{\beta}}
&
\begin{array}{c}
T_{\alpha' b\Sigma}\sim  T_{\alpha\Upsilon\Sigma}\sim T_{a\Upsilon}{}^\alpha\\
T_{ab'}{}^\beta\sim T_{\alpha b'\Sigma}\sim T_{\alpha'\Upsilon\Sigma}\\
\sim 
T_{\Upsilon\Upsilon}{}^\beta\sim T_{a\digamma}{}^{\beta'}\sim T_{\digamma\digamma}{}^{\beta}
\end{array}
\\\hline
2&
T_{\alpha}{}^{\beta'}{}_{\Sigma'},T_{\Upsilon}{}^{\alpha\beta},
T_{a\Upsilon\Sigma},T_{\digamma\digamma\Sigma}
&
\begin{array}{c}
T_{ab\Sigma}\sim T_{\alpha}{}^\beta{}_{\Sigma}\sim T_{a}{}^{\alpha\beta}\\
T_{a'b'\Sigma}\sim T_{\alpha}{}^\beta{}_{\Sigma'}\sim T_{\Upsilon}{}^{\alpha\beta'}\\
T_{a}{}^{\alpha\beta'}\sim T_{\alpha}{}^{\beta'}{}_{\Sigma}\sim T_{a'\Upsilon\digamma}\\
T_{ab'\Sigma}\sim T_{a}{}^{\alpha'\beta'}\\
T_{a\digamma\Sigma}\sim T_{\digamma}{}^{\alpha\beta}\\
 T_{\digamma}{}^{\alpha\beta'},T_{\Upsilon\Upsilon\Sigma},T_{\Upsilon\digamma\Sigma}
\end{array}
\\\hline
\end{array}}\label{torcon}
\eea 
The same relation holds for primed indices.
This formalism allows a simple treatment 
where all torsions with dimension 1 and less are structure constants 
otherwise zero.
\par
\vskip 6mm

\subsection{Vielbein}

Vectors in the space 
$\hat{\Lambda}_i=\Lambda_i{}^{\underline{M}}\mathring{\dd}_{\underline{M}}$
satisfy the following supersymmetric Lie bracket in the manifestly T-dual formulation
which is obtained from zero-mode of \bref{tratraf}
\bea
\left[\hat{\Lambda}_1,\hat{\Lambda}_2\right]_{\rm T}&=&-i\hat{\Lambda}_{12}\label{Tbra}
\\
\Lambda_{12}{}^{\underline{M}}&=&
\Lambda_{[1}{}^{\underline{N}}(D_{\underline{N}}\Lambda_{2]}{}^{\underline{M}})
-\frac{1}{2}\Lambda_{[1}{}^{\underline{N}}(D^{\underline{M}} \Lambda_{2]\underline{N}})
+\Lambda_1{}^{\underline{N}}\Lambda_2{}^{\underline{L}}f_{\underline{NL}}{}^{\underline{M}}
+k\Lambda_{(1}{}^{\underline{N}}(D^{\underline{M}} \Lambda_{2)\underline{N}})\nn
\eea
with arbitrary number $k$ which arises from an ambiguity of $\partial_\sigma\delta(\sigma)$.
The vielbein field $\hat{E}_{\underline{A}}=E_{\underline{A}}{}^{\underline{M}}\mathring{\dd}_{\underline{M}}$
has gauge symmetry generated by the manifestly T-dual bracket 
given in \bref{Tbra} with $k=-1/2$.
\bea
\delta_\Lambda\hat{E}_{\underline{A}} &=&i \left[\hat{E}_{\underline{A}},\hat{\Lambda}\right]_{\rm T}\nn\\
\delta_\Lambda E_{\underline{A}}{}_{\underline{M}}
&=&
D_{\underline{A}}\Lambda_{\underline{M}}-\Lambda^{\underline{N}}(D_{\underline{N}}E_{\underline{A}\underline{M}})
-E_{\underline{A}}{}^{\underline{N}}(D_{\underline{M}} \Lambda_{\underline{N}})
+E_{\underline{A}}{}^{\underline{N}}\Lambda^{\underline{L}}f_{\underline{NLM}}
\eea
with $\Lambda_{\underline{M}}=\Lambda^{\underline{N}}\eta_{\underline{NM}}$.

In order to identify components of the supervielbein field 
with the field content of type II supergravity,
dimensional reduction of two Lorentz degrees of freedom into 
one local Lorentz degree of freedom is realized as $S_{mn}+(-S_{m'n'})$ $\to$ 
$S_{\rm m n}$. 
Momentum/winding coordinates, 
$p_{{\rm m}} /\partial_\sigma x^{\rm m}$, related to
the left/right coordinates $P_{m},P_{m'}$  as 
$V_{\rm m}=(V_m+V_{m'})/2$ and $V^{\rm m}=(V_m-V_{m'})/2$.
The vielbein 
$E_{\underline{\rm a}}{}^{\underline{\rm m}}$  
 has the following form 
\bea
&&
E_{\underline{{\rm a}}}{}^{\underline{{\rm m}}}=
{\renewcommand{\arraystretch}{1.4}\left(\begin{array}{cc}
E_{\rm a}{}^{\rm m}&E_{\rm a}{}_{\rm m}\\
E^{\rm a}{}^{\rm m}&E^{\rm a}{}_{\rm m}
\end{array}\right)=
\left(\begin{array}{cc}
e_{\rm a}{}^{\rm m}&e_{\rm a}{}^{\rm n}B_{\rm nm}\\
0&e_{\rm m}{}^{\rm a}
\end{array}\right)}~\label{metB}\\
&&\Rightarrow~
E_{\underline{{\rm a}}}{}^{\underline{\rm m}}\eta_{\underline{\rm mn}}E_{\underline{\rm b}}{}^{\underline{\rm n}}
=\eta_{\underline{\rm ab}}~~,~~
E_{\underline{{\rm a}}}{}^{\underline{\rm m}}\eta^{\underline{\rm ab}}E_{\underline{\rm b}}{}^{\underline{\rm n}}
=\eta^{\underline{\rm mn}}~~,~~
\eta_{\underline{\rm ab}}=\eta_{\underline{\rm mn}}=
{\renewcommand{\arraystretch}{1.4}
\left(\begin{array}{cc}
0&{\bf 1}\\
{\bf 1}&0
\end{array}\right)}\nn\\
&&~~~E_{\underline{{\rm a}}}{}^{\underline{\rm m}}\hat{\eta}
^{\underline{\rm ab}}E_{\underline{\rm b}}{}^{\underline{\rm n}}
={\renewcommand{\arraystretch}{1.4}
\left(\begin{array}{cc}
G^{\rm mn}&G^{\rm ml}B_{\rm ln}\\
B_{\rm ml}G^{\rm lb}&G_{\rm mn}-B_{\rm ml}G^{\rm lk}B_{\rm kn}
\end{array}\right)~~,~~
\hat{\eta}^{\underline{\rm ab}}=\left(\begin{array}{cc}
{\bf 1}&0\\
0&{\bf 1}
\end{array}\right)}
\nn~~.
\eea  
The general coordinate transformations of $e_{\rm a}{}^{\rm m}$ and 
$B_{\rm mn}$ are given by $\delta_\xi \hat{E}=i[\hat{E},
\xi^{\underline{m}} \mathring{\dd}_{\underline{m}}]_{\rm T}$ 
as
\bea
\delta_\xi e_{\rm a}{}^{\rm m}&=&D_{\rm a}\xi^{\rm m}
-\xi^{\underline{n}}\partial_{\underline{n}} e_{\rm a}{}^{\rm m}
-e_{\rm a}{}^{\underline {\rm n}}\partial^{\rm m}\xi_{\underline {\rm n}}
+\omega_{\rm a}{}^{\rm ml}\xi_{\rm l}\nn\\
\delta_\xi B_{\rm mn}&=&\partial_{\rm [m}\xi_{\rm n]}
-\xi^{\underline{l}}\partial_{\underline{l}} B_{\rm mn}
-e_{[\rm m}{}^{\underline {\rm l}}\partial_{\rm n]}\xi_{\underline {\rm l}}
\eea
The gauge parameters are $\xi^{\underline{\rm m}}=(\xi^{\rm m},\xi_{\rm m})$ where 
$\xi_{\rm m}$ is the $B$ field gauge parameter.
The NS-NS field strength is 
$\omega_{\rm abc}=F_{\rm NS;}{}_{\rm abc}=\frac{1}{2}\partial_{\rm[a}  B_{\rm bc]}$, 
while $\omega^{\rm a}{}_{\rm bc}\sim e(\partial e)e$ is the usual part.
Gravitino and dilatino fields are included in $E_{a\Omega}$
as $e_{\rm m}{}^{\rm a}E_{\rm a}{}^{\mu}
=\psi_{\rm m}{}^{\mu}+(\gamma_{\rm m}\lambda)^\mu$.

 The R-R gauge field is
$e_{\rm m}{}^{\rm a}E_{{\rm a}\digamma}=C_{\rm RR;}{}_{\rm m}{}^{\mu\nu'}$,
where  two spinor indices $C^{\mu\nu'}$ are expanded by odd/even number 
antisymmetric gamma matrices
for type IIB/IIA.
The gauge transformation of the R-R gauge field
 is generated by 
$\delta_\Lambda \hat{E}=i[\hat{E},
\Lambda^{\mu\nu'} \mathring{\dd}_{\mu\nu'}]_{\rm T}$  
with $\Lambda_\digamma=\Lambda^\Upsilon=\Lambda^{\mu\nu'}$
as
\bea
&&\Lambda_\digamma=\Lambda^{\mu\nu'}
=\displaystyle\sum_{{\rm p= odd}}\Lambda_{\rm n_1\cdots n_p}
(\tau_{i})_{12}(\gamma^{\rm n_1\cdots n_p}
)^{\alpha\beta} ~{\rm with}~
{\renewcommand{\arraystretch}{1.2}\left\{\begin{array}{l}
\tau_i=\tau_1 ~{\rm for}~ {\rm p}=1,5 \\ 
\tau_i=i\tau_{2}~{\rm for}~{\rm p}=3
\end{array}\right.}\nn\\
&&e_{\rm m}{}^{\rm a}E_{a\digamma}=C_{\rm RR;}{}_{\rm m}{}^{\mu\nu'} =
\displaystyle\sum_{\rm p= even}(C_{\rm RR}){}_{\rm mn_1\cdots n_p}
(\tau_{i}){}_{12}(\rm \gamma^{\rm n_1\cdots n_p})^{\mu\nu}\nn\\
&&\delta{}C_{{\rm RR}}{}_{\rm m}^{\mu\nu'}
=\partial_{\rm m}\Lambda^{\mu\nu'}
~\Rightarrow~
\delta{}C_{{\rm RR}}{}_{\rm mn_1\cdots n_p}=
\partial_{[\rm m}\Lambda_{\rm n_1\cdots n_p]}
~~~.
\eea

Local supersymmetry transformations are generated by 
$\delta_\varepsilon \hat{E}=i[\hat{E},
\varepsilon^{\underline{\mu}} \mathring{\dd}_{\underline{\mu}}]_{\rm T}$  
with parameters 
$\varepsilon^{\underline{\mu}}=(\varepsilon^\mu,\varepsilon^{\mu'})$
in the first order of fermionic fields as 
\bea
\delta_\varepsilon e_{\rm a}{}^{\rm m}&=&
-B_{{\rm a}\underline{\mu}}\partial^{\rm m}\varepsilon^{\underline{\mu}}
+\psi_{\rm a}{}^{\underline{\mu}}
(\gamma^{\rm m}\varepsilon)_{\underline{\mu}}
\nn\\
\delta_\varepsilon B_{\rm mn}&=&
-B_{[{\rm m}|\underline{\mu}}\partial_{|{\rm n}]}
\varepsilon^{\underline{\mu}}
+\psi_{[{\rm m}}{}^{{\mu}}(\gamma_{\rm n]}\varepsilon)_{{\mu}}
-\psi_{[{\rm m}}{}^{{\mu}'}(\gamma_{\rm n]}\varepsilon)_{{\mu}'}
\nn\\
\delta_\varepsilon{}C_{{\rm RR}}{}_{\rm m}{}^{\mu\nu'}
&=&\frac{1}{2}\psi_{\rm m}{}^{(\mu}\varepsilon^{\nu')}
+\frac{1}{2}(\lambda \gamma_{\rm m})^{(\mu}\varepsilon^{\nu')}
-\frac{1}{2} (E_\digamma{}^\mu(\gamma_{\rm m}\varepsilon)_\mu 
-E_\digamma{}^{\mu'}(\gamma_{\rm m}\varepsilon)_{\mu'} )\nn\\
\delta_\varepsilon 
\psi_{\rm m}{}^\mu&=&\frac{1}{2}\left(
D_{\rm m}\varepsilon^\mu
+\omega_{\rm m}{}^{ nl}(\gamma_{ nl}\varepsilon)^\mu
-E^\mu{}_{\underline{\nu}}
\partial_{\rm m}\varepsilon^{\underline{\nu}}
+F_{\rm NS}{}^{\mu\nu}
(\gamma_{\rm m}\varepsilon)_\nu
+F_{\rm RR}{}^{\mu\nu'}
(\gamma_{\rm m}\varepsilon)_{\nu'}\right)
\nn\\
\delta_\varepsilon\psi_{\rm m}{}^{\mu'}&=&\frac{1}{2}\left(
D_{\rm m}\varepsilon^{\mu'}
-\omega_{\rm m}{}^{n'l'}(\gamma_{n'l'}\varepsilon)^{\mu'}
-E^{\mu'}{}_{\underline{\nu}}
\partial_{\rm m}\varepsilon^{\underline{\nu}}
+F_{\rm NS}{}^{\mu'\nu'}
(\gamma_{\rm m}\varepsilon)_{\nu'}
+F_{\rm RR}{}^{\mu'\nu}
(\gamma_{\rm m}\varepsilon)_\nu\right)\nn\\
\delta_\varepsilon E_{\digamma\digamma}&=&E_{\digamma}{}^{(\mu}\varepsilon^{\nu')}
\nn\\
\delta_\varepsilon E_{\digamma}{}^\mu
&=&\delta_\varepsilon E^{\nu\nu';\mu}~=~
C_{\rm RR}{}_{\underline{m}}{}^{\nu\nu'}\partial^{\underline{m}}\varepsilon^\mu
+E_{\digamma}{}^{mn}(\gamma_{mn}\varepsilon)^\mu
-E^{\mu(\nu}\varepsilon^{\nu')}
\nn
\eea
It is noted that spinor indices are denoted by $^{\mu,\nu,\cdots}$,
since our flat Lorentz indices are denoted by $S_{mn}$ and $S_{m'n'}$.
But later $^{\alpha,\beta,\cdots}$ is used 
after gauge fixing $S_{mn}=S_{ab}$ and $S_{m'n'}=S_{a'b'}$.
From the supersymmetry transformation of $C_{\rm RRm}{}^{\mu\nu'}$,
$E_{\digamma \Omega}$ turns out to include dilatino $\lambda$.
It is also derived that $E_{\digamma\digamma}$ 
include dilaton $\phi$ from its supersymmetry 
transformation rule.
Unlike the case of type II superspace without $\Upsilon_{\mu\nu'}$,
the dilaton and the dilatino are included in vielbein field without introducing 
$\hat{T}$ constraint which is obtained from the integral measure factor. 

From the supersymmetry transformation rule of the gravitinos 
 $E_{\underline{\Omega\Omega}}$ are identified with 
 NS-NS and R-R gauge field strengths,
$E_{\Omega\Omega}=E_{\rm NS}{}^{\mu\nu}$ and 
$E_{\Omega\Omega'}=E_{\rm RR}{}^{\mu\nu'}$
as same as before.
In addition to this 
the R-R gauge field strengths are included in
$E_{\digamma\Sigma}=F_{\rm RR}{}^{\alpha\beta' cd}$
from the supersymmetry transformation of the dilatino in $E_{\digamma \Omega}$.
The R-R field strength $F_{\rm RR}$ appears in $\{\dd_{\alpha},\dd_{\beta'}\}$,
so $T_{\alpha\beta'\Sigma}$
and $T_{a\alpha}{}^{\beta'}$ are related by the Bianchi identity.
These torsions were not zero without central extension,
but they are absorbed by vielbein
as $T_{\alpha\beta'\Sigma}=f_{\alpha\beta\digamma}E_{\Upsilon\Sigma}$
and $T_{a\alpha}{}^{\beta'}=f_{a \alpha \gamma}E^{\gamma \beta'}$.

As a result vielbein fields are identified  to supergravity multiplets  
from Bianchi identities, torsion constraints and gauge transformations as
follows:
\bea
{\renewcommand{\arraystretch}{1.8}
\begin{array}{l|l}
{\rm type ~II~string ~field}&{\rm fields}~({\rm dim.})\\
\hline
{\rm gravitational~metric}&e_{(\underline{am})}=E_{(\underline{am})},~
e_{\underline{\alpha}}{}^{\underline{\mu}}
=E_{\underline{\alpha}\Omega}(0)\\
B~{\rm fields}&B_{\underline{\alpha\mu}}=E_{\underline{\alpha\mu}}(-1),~
B_{\underline{a\mu}}=E_{\underline{a\mu}}(-\frac{1}{2}),~
B_{\underline{am}}=E_{[\underline{am}]}(0)\\
C~{\rm fields}&C_{\underline{\alpha}}{}^{\mu\nu'}
=E_{\underline{\alpha}\digamma}(-\frac{1}{2}),~
C^{\rm RR}{}_{\underline{a}}{}^{\mu\nu'}
=E_{\underline{a}\digamma}(0)\\
{\rm dilaton}&\varphi=E_{\digamma\digamma}(0)\\
{\rm dilatino}&\lambda_{\underline{\mu}}=E_{\digamma\Omega},~
E_{\underline{a}\Omega}(\frac{1}{2})\\
{\rm gravitino}&\psi_{\underline{a}}{}^{\underline{\mu}}
=E_{\underline{a}\Omega}(\frac{1}{2})\\
{\rm superconnections}&\omega_{\underline{\alpha}}{}^{\underline{mn}}
=E_{\underline{\alpha}\Sigma}(\frac{1}{2}),~
\omega_{\underline{\alpha}}{}^{\underline{mn}}
=E_{\underline{a}\Sigma}(1),~
\omega^{\underline{\alpha}}{}^{\underline{mn}}
=E_{\Omega\Sigma}(\frac{3}{2})\\
{\rm NS-}{\rm NS~field~strength}&
F_{\rm NS}{}^{[\mu\nu]},F_{\rm NS}{}^{[\mu'\nu']}=
E_{[\Omega\Omega]},E_{[\underline{a}\Sigma]}(1)\\
{\rm R-}{\rm R~field~strength}&
F_{\rm RR}{}^{\mu\nu'}=
E_{\Omega\Omega'},E_{\digamma\Sigma}(1)\\
{\rm covariance~compensator}&r^{\underline{abmn}}=E_{\Sigma\Sigma}(2)
\end{array}
}\label{vievie}
\eea
\par\vskip 6mm

\subsection{Linearized solution}
The vielbein in a weak field expansion with lower indices, 
$E_{\underline{AM}}=E_{\underline{A}}{}^{\underline{N}}\eta_{\underline{NM}}$, 
are written 
as 
\bea
&&E_{\underline{AM}}=\eta_{\underline{AM}}
+H_{\underline{A}}{}^{\underline{B}}\eta_{\underline{BM}}~~\Rightarrow~~
H_{\underline{AB}}=
H_{\underline{A}}{}^{\underline{C}}\eta_{\underline{CB}}
=\frac{1}{2}H_{[\underline{AB})}\nn~~~.
\eea
Coordinates corresponding to 
Lorentz symmetries and the R-R central charge are auxiliary coordinates, 
so it is possible to choose 
$E_{\underline{SM}}=\eta_{\underline{SM}}$ and 
$E_{\Upsilon{\underline{M}}}=\eta_{\Upsilon{\underline{M}}}$.
The vielbein fields given by
\bref{vievie} in linearized level are also denoted as
\bea
&&~~~~~~~~~~~~~~~~~~~~~~~~~D_{\underline{\beta}}~~~~~~~~~P_{\underline{b}}~~~~~~~~~~~~~\digamma^{\beta\beta'}~~~~~~~~~~~
\Omega^{\underline{\beta}}~~~~~~~~~~~\Sigma^{\underline{bd}}\nn\\
&&H_{\underline{AB}}=
{\renewcommand{\arraystretch}{1.8}\begin{array}{c}
D_{\underline{\alpha}}\\
P_{\underline{a}}\\
\digamma^{\alpha\alpha'}\\
\Omega^{\underline{\alpha}}\\
\Sigma^{\underline{ac}}\end{array}
\left(\begin{array}{ccccc}
B_{\underline{\alpha \beta}}&-B_{\underline{a \alpha }}&
-C_{\underline{\alpha}}{}^{\beta \beta'}&
h_{\underline{\alpha}}{}^{\underline{\beta}}&
\omega_{\underline{\alpha}}{}^{\underline{bd}}\\
B_{\underline{a\beta}}&h_{\underline{ab}},B_{\underline{ab}}&C_{\rm RR}
{}_{\underline{a}}{}^{\beta\beta'}&
\psi_{\underline{a}}{}^{\underline{\beta}},\lambda&
\omega_{\underline{a}}{}^{\underline{bd}},F_{\rm NS}\\
C_{\underline{\beta}}{}^{\alpha \alpha'}&
-C_{\rm RR}{}_{\underline{b}}^{\alpha\alpha'} &\phi&\lambda
&F_{\rm RR}{}^{\alpha\alpha'\underline{bd} }\\
h^{\underline{\alpha}}{}_{\underline{\beta}}&
-\psi_{\underline{b}}{}^{\underline{\alpha}},\lambda
&\lambda&F_{\rm NS}^{\underline{\alpha\beta}},F_{\rm RR}^{{\alpha\beta'}}
&\omega^{\underline{\alpha bd}}\\
-\omega_{\underline{\beta}}{}^{\underline{ac}}&-\omega_{\underline{b}}{}^{\underline{ac}},F_{\rm NS}&
-F_{\rm RR}{}^{\beta\beta' \underline{ac}}
&-\omega^{\underline{\beta ac}}&r^{\underline{acbd}}
\end{array}
\right)}\label{vielbein}
\eea
Torsions  in linearized level are given by
\bea
T_{\underline{ABC}}&=&
\frac{1}{2}\left(D_{[\underline{A}}H_{\underline{BC}])}
-H_{[\underline{A}}{}^{\underline{D}}f_{\underline{BC]D}}\right)
\eea
where $D_{\underline{A}}$ is a flat space covariant derivative.
The generalized gauge transformations are
\bea
\delta_\Lambda H_{\underline{AB}}&=&
D_{[\underline{A}}\Lambda_{\underline{B})}
-\Lambda^{\underline{C}}f_{\underline{CAB}}~~~.\label{gaugelinear}
\eea

Let us solve torsion constraints given by \bref{torcon} in the linearized level.
Although there are new torsions appeared by  the central extension 
with indices $\Upsilon$ and $\digamma$,
many torsion constraints  whose indices do not include  $\Upsilon$ and $\digamma$  are
 the same as the previous paper \cite{Hatsuda:2014qqa} as;
 \bea
T_{\underline{ABC}}&=&\frac{1}{2}\left(D_{[\underline{A}}H_{\underline{BC}]}
+H_{[\underline{A}}{}^{\underline{D}}f_{\underline{D}|\underline{BC}]}
+H_{[\underline{A}|{\Upsilon}}f_{{\digamma}|\underline{BC}]}
+H_{[\underline{A}|{\digamma}}f_{{\Upsilon}|\underline{BC}]}\right)\nn\\
&=&\frac{1}{2}\left(D_{[\underline{A}}H_{\underline{BC}]}
+H_{[\underline{A}}{}^{\underline{D}}f_{\underline{D}|\underline{BC}]}
\right)\nn~~,~
_{\underline{A}\neq \Upsilon,\digamma}\eea
because of  $H_{\Upsilon\underline{A}}=f_{\Upsilon\underline{AB}}=0$.
Torsions with two $\Upsilon$ indices are zero because of the same reason
\bea
T_{\underline{A}\Upsilon\Upsilon}&=&\frac{1}{2}\left(D_{[\underline{A}}H_{\underline{\Upsilon\Upsilon}]}
+H_{[\underline{A}}{}^{\underline{D}}f_{\underline{D}|\underline{\Upsilon\Upsilon}]}
\right)~=~0\nn~~~.
\eea
Therefore nontrivial new torsions are
\bea
T_{\Upsilon\underline{AB}},~ T_{\Upsilon\digamma\underline{A}},~
T_{\Upsilon\digamma\digamma},~
T_{\digamma\underline{AB}},~
T_{\digamma\digamma\underline{A}},~ 
T_{\digamma\digamma\digamma}~~{\rm for}~
_{\underline{A},\underline{B}\neq \Upsilon,\digamma}
\eea
New torsion constraints involving $\Upsilon$ and $\digamma$ indices 
are solved in linearized level:
\bea
{\renewcommand{\arraystretch}{1.6}
\begin{array}{lcl}
{\rm dim}&{\rm torsion~constraints}&\Rightarrow~{\rm field} \\\hline
0&T_{\alpha\underline{\beta}\Upsilon}=0&D_\Upsilon B_{\alpha\underline{\beta}}=0\\
&T_{\alpha\beta\digamma}=0&
C^{\rm RR}{}_{a}{}^{\gamma\gamma'}=-D_{(\alpha} C_{\beta)}{}^{\gamma\gamma'}(\gamma_a)^{\alpha\beta}
-h_\alpha{}^{\gamma'}(\gamma_a)^{\alpha\gamma}\\
&T_{\alpha\beta'\digamma}=f_{\alpha\beta'\digamma}&
h_\alpha{}^\gamma f_{\beta'\gamma\digamma}-h_{\beta'}{}^{\gamma'}f_{\alpha\gamma'\digamma}
=D_{(\alpha} C_{\beta')\digamma}\\\hline
\frac{1}{2}&T_{\alpha\underline{b}\Upsilon}=0&D_\Upsilon B_{\alpha\underline{b}}=0\\
&T_{\alpha\Upsilon\digamma}=0&\omega_{\alpha\Sigma}=-D_{\Upsilon}C_{\alpha\digamma}\\
&T_{\alpha a\digamma}=0&\psi_a{}^{\beta'}f_{{\alpha}\beta'\digamma}
+H_{\digamma}{}^\beta(\gamma_a)_{\alpha\beta}=
D_\alpha C^{\rm RR}{}_{a\digamma}-D_aC_{\alpha\digamma}\\
&T_{\alpha a'\digamma}=0&\psi_{a'}{}^{\beta'}f_{\alpha\beta'\digamma}=
-D_\alpha C^{\rm RR}{}_{a'\digamma}+D_{a'}C_{\alpha\digamma}\\
&T_{\alpha\digamma\digamma}=0&H_{[\digamma|}{}^{\beta'}f_{\alpha\beta'|\digamma]}
=D_\alpha H_{\digamma\digamma}\\\hline
1&T_{\alpha\Upsilon}{}^{\underline{\beta}}=0&
D_\Upsilon h^{\underline{\beta}{}_{\alpha}}=0\\
&T_{a\underline{b}\Upsilon}=0&
D_\Upsilon h_{a\underline{b}}=0\\
&T_{a\Upsilon\digamma}=0&\omega_{a\Sigma}=-D_{\Upsilon}C^{\rm RR}{}_{a\digamma}\\
&T_{\Upsilon\digamma\digamma}=0&F_{\rm RR}{}_{\digamma\Sigma}=D_\Upsilon H_{\digamma\digamma}\\
&T_{\alpha\digamma}{}^\beta=0&F_{\rm RR}{}_{\digamma}{}^{ab}(\gamma_{ab})^\beta{}_\alpha
-F_{\rm RR}{}^{\beta\beta'}f_{\alpha\beta'\digamma}=-D_\alpha H_\digamma{}^\beta\\
&T_{\alpha\digamma}{}^{\beta'}=0&F_{\rm NS}{}^{\alpha\beta'}f_{\alpha\beta'\digamma}=-D_\alpha
H_\digamma{}^{\beta'}\\
&T_{ab\digamma}=0&F_{\rm RR}{}_{\digamma ab}=-D_{[a}C^{\rm RR}{}_{b]\digamma}\\
&T_{ab'\digamma}=0&D_{[a}C^{\rm RR}{}_{b']\digamma}=0\\
&T_{a\digamma\digamma}=0&D_aH_{\digamma\digamma}=0\\\hline
3/2&T_{a\Upsilon}{}^{\beta'}=0&D_\Upsilon\psi_a{}^{\beta'}=0\\
&T_{\Upsilon\digamma}{}^{\beta}=0&\omega^\beta{}_\Sigma f_{S\Upsilon\digamma}=
D_\Upsilon H_\digamma{}^\beta\\
&T_{a\digamma}{}^\beta=0&D_aH_{\digamma}{}^\beta=0\\
&T_{\alpha\digamma\Sigma}=0&\omega^{\beta'ab}=D_\alpha F_{\rm RR}{}^{\alpha\beta'ab}\\\hline
2&T_{\Upsilon}{}^{\alpha\beta}=0&D_\Upsilon F_{\rm NS}{}^{\alpha\beta}=0\\
&T_{a\Upsilon\Sigma}=0&D_\Upsilon \omega_{a}{}^{bc}=0\\\hline
\end{array}}
\eea
where we assumed $D_{\Sigma}E=D_{\Omega}E=D_{\digamma}E=0$ as
a simple expression although it is not necessary.

\par\vskip 6mm
\section{Green-Schwarz superstring action}

In this section we write down actions for strings in the manifestly T-duality formalism.
The Wess-Zumino terms are written by
 bilinear combination of left-invariant local currents.
Equations of motion are chirally separated.

\subsection{Action}

For a graded Lie algebra $G_{\underline{M}}$ 
\bea
\lbrack G_{\underline{M}},G_{\underline{N}}\}
=if_{\underline{MN}}{}^{\underline{K}}G_{\underline{K}}~~~,
\eea
the left-invariant current $J$ is   written as
\bea
g^{-1}(dg)={J}&,&dJ=-J\wedge J\nn\\
J=iJ^{\underline{M}}G_{\underline{M}}&,&dJ^{\underline{M}}
=-\frac{1}{2}J^{\underline{K}}
\wedge J^{\underline{N}}f_{\underline{NK}}{}^{\underline{M}}~
~~.\nn
\eea
It is related to the covariant derivative as
\bea
g^{-1}(dg)
=idZ^{\underline{M}}E_{\underline{M}}{}^{\underline{A}}G_{\underline{A}}
&,&\nabla_{\underline{A}}=\frac{1}{i}E_{\underline{A}}{}^{\underline{M}}\partial_{\underline{M}}~
.
\eea
It is convenient to classify the left-invariant currents 
 by the canonical dimensions of generators 
 denoted by $s$, where they satisfy following
Maurer-Cartan (MC) equations for left-invariant currents
\bea
&{J}= \displaystyle\sum_{s=0}^{2} J_{[s]}~~,
~~dJ_{[s]}=-\displaystyle\sum_{t=0}^{s} J_{[s-t]}\wedge J_{[t]}&.
\eea
Right-invariant currents are decomposed as follows and they satisfy the following
 MC equations for $j$
\bea
&(dg)g^{-1}= {j}~~,~~dj=j\wedge j&\nn\\
&j_{[s]}\equiv gJ_{[s]}g^{-1}~~,~~j= \displaystyle\sum_{s=0}^{2} j_{[s]}~~,
~~dj_{[s]}=j\wedge j_{[s]}+ j_{[s]}\wedge j-
\displaystyle\sum_{t=0}^{s} j_{[s-t]}\wedge j_{[t]}~~.&
\eea
It is noted that 
 $j_{[s]}$ does not have a definite dimension $s$,
 but it may contain several generators with different dimension $s$.

We introduce three kinds of traces;
\bea
{\renewcommand{\arraystretch}{1.8}
\begin{array}{lll}
{\rm trace}&{\rm rank}&{\rm reference~eq.}\\
{\rm tr}\Bigl(G_{\underline{M}}G_{\underline{N}}\Bigr)=\eta_{\underline{MN}}
&{\rm nondegenerate},~
\eta_{\underline{MN}}\eta^{\underline{NL}}=\delta_{\underline{M}}^{\underline{L}}
& \bref{etaMNRR}
\nn\\
\hat{\rm tr}\Bigl(G_{\underline{M}}G_{\underline{N}}\Bigr)\equiv
\hat{\eta}_{\underline{MN}}
&{\rm not~always~nondegenerate}& \bref{etahat}\nn\\
{\rm tr}_B\Bigl(G_{\underline{M}}G_{\underline{N}}\Bigr) \equiv
{\eta}_{\underline{MN}} |_{
\underline{M}\in \tilde{I},
\underline{N}\in {I}}&{\rm almost~half} -{\rm rank}&\bref{bilinear}
\end{array}}
\nn
\eea
Three kinds of ``traces" are non-vanishing only if
${\rm dim}(G_{\underline{M}})+{\rm dim}(G_{\underline{N}})=2$.
Indices $_{I}$ and $_{\tilde{I}}$ are classified in \bref{bilinear}.
Cyclicity holds  for tr($gh$)=tr($hg$).
The structure constant with lower indices is
 totally graded-antisymmetric 
\bea
&&{\rm tr}\left([G_{\underline{M}},G_{\underline{N}}\}G_{\underline{K}}\right)=if_{\underline{MNK}}~~~,\nn\eea
 which  are nonvanishing only if dim
 $(G_{\underline{M}})+{\rm dim}(G_{\underline{N}})+ {\rm dim}(G_{\underline{K}})=2$.

We propose the following Green-Schwarz action 
in manifestly T-duality formalism,
\bea
I&=&\displaystyle\int~d\tau d\sigma~[{\cal L}_0+{\cal L}_{\rm WZ}] \label{WZBeta}\\
{\cal L}_0&=&
\frac{1}{2}J^{\underline{M}}_{[1]}\wedge\ast J^{\underline{N}}_{[1]}~
\hat{\eta}_{\underline{MN}}
~=~-\frac{1}{2}\hat{\rm tr}~ J_{[1]}\wedge \ast J_{[1]}\nn\\
{\cal L}_{\rm WZ}&=&
\frac{\cal K}{2}J^{\underline{M}}\wedge J^{\underline{N}}B_{\underline{MN}}~=~
{\cal K}\sum_{s=0}^{1}
J_{[2-s]}^{\tilde{J}}\wedge J_{[s]}^{I}~\eta_{\tilde{I}{J}}
=-{\cal K}\sum_{s=0}^{1}{\rm tr}_B ~
J_{[2-s]}\wedge J_{[s]}\nn~~~.
\eea
with ${\cal K}=\pm 1$.
The  $B_{\underline{MN}}$ fields in flat space are graded antisymmetric constants as 
\bea
B_{\underline{MN}}=\frac{1}{2}B_{[\underline{MN})}=
\left\{\begin{array}{ll}
\eta_{\underline{MN}}&\cdots _{\underline{M}\in \tilde{I},\underline{N}\in {I}}\\
-\eta_{\underline{MN}}&\cdots _{\underline{M}\in {I},\underline{N}\in \tilde{I}}\\
0&\cdots _{\underline{MN}\in {I}}
\end{array}
\right. ~~~.
\eea
Even for a bosonic string it is useful to include the Wess-Zumino term
with respect to fluxes corresponding to 
structure constants of the nondegenerate algebra.
Currents included in the Wess-Zumino term 
do not appear in the kinetic term ${\cal L}_0$,
so they are auxiliary degrees of freedom. 
Dimensional reduction constraints are imposed to remove these auxiliary coordinates.

Let us evaluate the exterior derivative of the Wess-Zumino term
;
\bea
d{\cal L}_{\rm WZ}&=&-d\Bigl({\cal K}\sum_{s=0}^{1}{\rm tr}_B ~
J_{[2-s]}\wedge J_{[s]}\Bigr)\nn\\&=
&-{\cal K}\sum_{s=0}^{1}{\rm tr}_B 
\left(dJ_{[2-s]}\wedge J_{[s]}-J_{[2-s]}\wedge dJ_{[s]}\right)\nn\\
&=&-
{\cal K}\sum_{s=0}^{1}{\rm tr}_B 
\Bigl(-\sum_{t=0}^{2-s}J_{[2-s-t]}\wedge J_{[t]}\wedge J_{[s]}
+\sum_{t=0}^{s}J_{[2-s]}\wedge J_{[s-t]}\wedge J_{[t]}\Bigr)\nn\\
&=&{\cal K}\Bigl[\frac{1}{3}{\rm tr} ~J\wedge J\wedge J+
{\rm tr}_B~\Bigl(\sum_{t=\frac{1}{2}}^{1}J_{[1]}\wedge 
 J_{[1-t]}\wedge J_{[t]}\Bigr)
\Bigr]\label{3form}~~~.
\eea
The Wess-Zumino term is   Chevalley-Eilenberg  cohomology trivial in the centrally extended algebra, 
while non-trivial in usual case. 
The usual Green-Schwarz superstring action has only second term 
in the right hand side of  \bref{3form},
where the Wess-Zumino term cannot be written down 
 in bilinear form of the currents locally in 2-dimensional worldsheet.
The first term in \bref{3form} gives all fluxes corresponding to
structure constants  of the global symmetry algebra.
It turns out that this term makes \bref{3form} exact three form and the 
 Wess-Zumino term is obtained in the bilinear of currents
 as in \bref{WZBeta}.

Under an arbitrary variation 
the left-invariant current becomes 
\bea
\delta J= \delta(g^{-1}d g)=g^{-1}(d \Delta)g~~,~~\Delta=\delta g g^{-1}~~~.
\eea
The variation of the first term of \bref{3form} is
total derivative written in terms of one form currents as
\bea
\delta ~\frac{1}{3}{\rm tr}~J\wedge J\wedge J&=&-
d\Bigl({\rm tr}~(\Delta~dj)\Bigr)~~~,
\eea  
which gives a contribution of current divergence to an equation of motion.

We take variation of the local form of the Wess-Zumino term 
in  \bref{WZBeta} directly:
\bea
\delta {\cal L}_{\rm WZ}
&=&
{\cal K}{\rm tr}_B~
\Delta \sum_{s=0}^{1} d( j_{[2-s]}-j_{[s]})\nn\\
&=&
{\cal K}\Bigl[
{\rm tr}\Delta( -dj+\sum_{0<t<2}\sum_{0<r<t}j_{[t-r]}\wedge j_{[r]})
+
{\rm tr}_B
\Delta \sum_{s>0}^{1} d( j_{[2-s]}-j_{[s]})\Bigr]
\eea
The variation of the kinetic term becomes
\bea
\delta {\cal L}_0&=&\hat{\rm tr}~\left[
\Delta~ d\ast j_{[1]}\right] ~~~.
\eea

\par\vskip 6mm

\subsection{T-dual bosonic string}

One form currents of a  bosonic string 
are denoted by
\bea
& J_{[0]}=\{S,~S'\}~~,~~
 J_{[1]}=\{P,~P'\}~~,~~
 J_{[2]}=\{\Sigma,~\Sigma'\}\nn~~~&\\
 & j_{[0]}=\{s,~s'\}~~,~~
 j_{[1]}=\{p,~p'\}~~,~~
 j_{[2]}=\{\sigma,~\sigma'\}\nn~~&
\nn\\
&j=s+p+\sigma~,~j'=s'+p'+\sigma'& 
\eea
For cases where two sets of algebras are decoupled, $J$ and $J'$,
$j$ and $j'$ are functions only on $Z^M$ and $Z^{M'}$
because of $g=g(Z^M)g(Z^{M'})=g(Z^{M'})g(Z^{M})$.
The left-invariant currents satisfy the following  MC equations  
\bea
&& g^{-1}dg=J+J'\nn\\
&&J=S+P+\Sigma,~J'=S'+P'+\Sigma'\nn\\
&&{\renewcommand{\arraystretch}{1.4}
\left\{\begin{array}{l}
d {S}= -{S}\wedge {S}\\
d {P}= -{P}\wedge {S}- {S}\wedge {P}\\
d {\Sigma}= -{\Sigma}\wedge {S}- {S}\wedge {\Sigma}
- {P}\wedge {P}\\
0=P\wedge \Sigma+\Sigma\wedge P=\Sigma\wedge\Sigma
\end{array}\right.},~
{\renewcommand{\arraystretch}{1.4}
\left\{\begin{array}{l}
d {S'}= -{S'}\wedge {S'}\\
d {P'}= -{P'}\wedge {S'}-{S'}\wedge {P'}\\
d {\Sigma'}=- {\Sigma'}\wedge {S'}- {S'}\wedge {\Sigma}'
-{P'}\wedge {P'}\\
0=
P'\wedge \Sigma'+\Sigma'\wedge P'=\Sigma'\wedge\Sigma'
\end{array}\right.}
\eea
Components of the right invariant current satisfy the following MC equations 
\bea
&& g^{-1}dg=j+j'\nn\\
&&j=s+p+\sigma,~j'=s'+p'+\sigma'\nn\\
&&{\renewcommand{\arraystretch}{1.4}
\left\{\begin{array}{l}
d {s}=j\wedge j- {p}\wedge {p}\\
d {p}=2 {p}\wedge  {p}\\
d {\sigma}= -{p}\wedge  {p}\\
0=p\wedge \sigma+\sigma\wedge p=\sigma\wedge\sigma
\end{array}\right.},~
{\renewcommand{\arraystretch}{1.4}
\left\{\begin{array}{l}
d {s'}=j'\wedge  j'- {p'}\wedge {p'}\\
d {p'}=2 {p'}\wedge  {p'}\\
d {\sigma'}= -{p'}\wedge  {p'}\\
0=p'\wedge \sigma'+\sigma'\wedge p'=\sigma'\wedge\sigma'
\end{array}\right.}\label{MCb}
\eea

An action for a bosonic string in the manifestly T-duality formalism is given by  
\bea
{\cal L}_0&=&-\frac{1}{2}
\hat{\rm tr}\Bigl[P\wedge* P+P'\wedge \ast {P'}\Bigr]~=~
-\frac{1}{2}
{\rm Tr}\Bigl[P\wedge* P+P'\wedge \ast {P'}\Bigr]
\nn\\
{\cal L}_{\rm WZ}&=&-{\cal K} {\rm tr}_B\Bigl[{\Sigma}\wedge {S} +
{\Sigma'}\wedge {S'}\Bigr]~=~-{\cal K} {\rm Tr}\Bigl[{\Sigma}\wedge {S} -
{\Sigma'}\wedge {S'}\Bigr]~~~
\eea
with ${\rm Tr}(G_{{M}}G_{{N}})=\eta_{{MN}}$ and 
 ${\rm Tr}(G_{{M'}}G_{{N'}})=|\eta_{{M'N'}}|$.
From the MC equation \bref{MCb} 
\bea
d( j_{[2]}-j_{[0]})~=~
d(\sigma-s+\sigma'-s')~=~-d(j+j')~~~,
\eea
the variation of the action is given by
\bea
\delta {\cal L}_0&=&\hat{\rm tr}~\left[
\Delta~ d\ast (p+p')\right]~=~{\rm Tr}~\left[
\Delta~ d\ast (p+p')\right]\nn\\
\delta {\cal L}_{\rm WZ}&=&-{\cal K}~{\rm tr}
\Bigl[\Delta~ d (j+j')\Bigr]~=~
-{\cal K}~{\rm Tr}\Bigl[
\Delta~ d (j-j')\Bigr]
\eea
with $\ast\ast=1$.
As a result the equation of motion is given by
\bea
&d\ast\Bigl(p-{\cal K}\ast(p+\sigma+s) \Bigr)=0
~,~d\ast\Bigl(p'+{\cal K}\ast(p'+\sigma'+s') \Bigr)=0&
\eea
or equivalently 
\bea
&\partial_i\Bigl(p^i 
-{\cal K}\epsilon^{ij}(p+\sigma+s)_j \Bigr)=0~,~
\partial_i\Bigl(p'{}^i 
+{\cal K}\epsilon^{ij}(p'+\sigma'+s')_j \Bigr)=0~~,~~
_{i,j=\tau,\sigma}.&\nn
\eea
Since left and right sets are decoupled, 
 the left and the right coordinates satisfy 
 equations of motion independently. 
The nondegenerate metric $\eta_{\underline{MN}}$
has opposite signatures for left and right coordinates,
so the coefficients of the Wess-Zumino term have 
opposite signs  for left and right sectors.
The equations of motions for left and right sectors become different
chirality
with
$\partial_\pm=\partial_\sigma\pm\partial_\tau$ and
$ {j}_\pm= {j}_\sigma\pm {j}_\tau$ 
as
\bea
{\cal K}=-1~&\Rightarrow&~
{\renewcommand{\arraystretch}{1.6}
\left\{\begin{array}{l}
\partial_+\left(p_-+\frac{1}{2}(s_-+{\sigma}_-)\right)
-\frac{1}{2}\partial_-(s_++\sigma_+)=0
\\
\partial_-\left(p'_++\frac{1}{2}(s'_++\sigma'_+)\right)
-\frac{1}{2}\partial_+(s'_-+\sigma'_-)=0
\end{array}\right.}\nn\\
{\rm or}&&\label{eqchiralsuperprime}~~~\\
{\cal K}=1~&\Rightarrow&~
{\renewcommand{\arraystretch}{1.6}
\left\{\begin{array}{ll}
\partial_-\left(p_++\frac{1}{2}(s_++{\sigma}_+)\right)
-\frac{1}{2}\partial_+(s_-+\sigma_-)=0
\\
\partial_+\left(p'_-+\frac{1}{2}(s'_-+\sigma'_-)\right)
-\frac{1}{2}\partial_-(s'_++\sigma'_+)=0
\end{array}\right.}
\nn
\eea
Since coordinates of Lorentz symmetry $S_{mn}$, $S_{m'n'}$ and their 
nondegenerate partners $\Sigma^{mn}$, $\Sigma^{m'n'}$ do not have 
kinetic terms, they are auxiliary variables.
For a coset space Poincar\'{e}/Lorentz, $S_{mn}=S_{m'n'}=0$ 
are coset constraints  and $\Sigma^{mn}$ and ${\Sigma}^{m'n'}$  
are also constrained as gauge fields \cite{Hatsuda:2005te}.
From the algebra \bref{SDPomesig} 
both $S^{mn}=0$ and $\Sigma^{mn}=0$ are not
first class constraints, but
$S^{mn}=0$  and the symmetry generator
 $\tilde{\Sigma}=0$ are first class constraints 
 where symmetry generators commute with  covariant derivatives. 
Then  currents $J_{[0]}=S,~S'$ and $j_{[2]}=\sigma,~\sigma'$ 
are gauged away as gauge fixing conditions.
Details are discussed in another paper \cite{HKS14}. 
Then left/right  currents $p_i$ and $p'_i$ are essentially 
chiral currents,
 e.g. $p_+=\partial_+ x$ and 
 $p'_-=\partial_- x'$ in flat space for ${\cal K}=1$.
Equation of motion for bosonic coordinates are
$\partial_-p_+=\partial_-\left(\partial_+ x\right)=0$ and $
\partial_+p_-=\partial_+\left(\partial_- x'\right)=0$.
The dimensional reduction condition reduces double coordinates 
into the usual one, for example $p-p'=0$ and $X=x+x'$,
then equations of motion reduce into the usual one,
$\partial_+\partial_- X=0$.

\par\vskip 6mm

\subsection{Type II superstring}

For a type II superstring  
one form currents of the nondegenerate  super-Poincar\'{e} algebra 
are denoted by
\bea
& J_{[0]}=\{S,~S'\}~,~J_{[\frac{1}{2}]}=\{D,~D'\}~,~
J_{[1]}=\{P,P'\}~,~
J_{[\frac{3}{2}]}=\{\Omega,\Omega' \}~,~
J_{[2]}=\{\Sigma,~\Sigma'\}\nn~~~&\\
 &j_{[0]}=\{s,~s'\}~,~j_{[\frac{1}{2}]}=\{{\rm d},~{\rm d}'\}~,~
j_{[1]}= \{p,~p'\}~,~
j_{[\frac{3}{2}]}=\{\omega,~\omega'\}~,~
j_{[2]}=\{\sigma,~\sigma'\}\nn~~.&
\eea
The left-invariant currents and the MC equations are 
\bea
g^{-1}dg&=& {J}+J'\nn\\
 {J}&=&
 {S}+ {D}+ {P}+ {\Omega}+ {\Sigma}
~,~J'~=~ {S}'+ {D}'+ {P}'+ {\Omega}'+ {\Sigma}'
 \nn\\
d {S}&=& -{S}\wedge  {S}\nn\\
d {D}&=& -{S}\wedge  {D}- {D}\wedge  {S}\nn\\
d {P}&=& -{S}\wedge  {P}- {P}\wedge  {S}- {D}\wedge  {D}\nn\\
d {\Omega}&=& -{S}\wedge  {\Omega}
-{\Omega}\wedge  {S}- {D}\wedge  {P}
- {P}\wedge  {D}
\nn\\
d {\Sigma}&=& -{S}\wedge  {\Sigma}
- {\Sigma}\wedge  {S}-{D}\wedge  {\Omega}
- {\Omega}\wedge  {D}-
 {P}\wedge  {P}~~~\\
 0&=& \Sigma\wedge\Sigma=\Sigma\wedge \Omega+\Omega\wedge \Sigma=\Sigma\wedge P+P\wedge \Sigma\nn\\
 &=&\Sigma\wedge D+D\wedge \Sigma=\Omega\wedge \Omega=\Omega\wedge P+P\wedge \Omega~~~.
 \nn
\eea
The MC equations for primed left-invariant currents are the same as above.
Decomposed right-invariant currents and the MC equations are
\bea
(dg)g^{-1}&=&{j}+j'~\nn\\
{j}&=&
 {s}+ {\rm d}+ {p}+ {\omega}+ {\sigma}
~,~j'~=~{s}'+ {\rm d}'+ {p}'+ {\omega}'+ {\sigma}'
 \nn\\
d {s}&=&j\wedge j-p\wedge p-{\rm d}\wedge {\rm d}
- {{\rm d}}\wedge (  {p}+ {\omega})
-({p}+ {\omega})\wedge {{\rm d}}
\nn\\
d {{\rm d}}&=&2 {{\rm d}}\wedge  {{\rm d}}
+ {{\rm d}}\wedge ( {p}+  {\omega})+( {p}+  {\omega})\wedge  {{\rm d}}
\nn\\
d {p}&=& -{{\rm d}}\wedge  {{\rm d}}+2 {p}\wedge  {p}
+ {{\rm d}}\wedge  {p}+ {p}\wedge  {{\rm d}}\nn\\
d {\omega}&=& {{\rm d}}\wedge(-  {p}+ {\omega})
+(  -{p}+ {\omega})\wedge {{\rm d}}\nn\\
d {\sigma}&=& -{p}\wedge  {p}
- {{\rm d}}\wedge  {\omega}- {\omega}\wedge  {{\rm d}}~\nn\\
0&=&\sigma\wedge\sigma=\sigma\wedge\omega+\omega\wedge\sigma=
\sigma\wedge p+p\wedge\sigma\nn\\
&=&\sigma\wedge {\rm d}+{\rm d}\wedge\sigma=\omega\wedge\omega
=\omega\wedge p+p \wedge\omega 
~~~.\label{mcsuper0}
\eea
The MC equations for primed ones are the same as the above with replacing unprimed indices with primed indices.

An action for a superstring in the manifestly T-duality formalism is given by
\bea
{\cal L}_0&=&-\frac{1}{2}
\hat{\rm tr}\Bigl[P\wedge* P+P'\wedge \ast {P'}\Bigr]
~=~-\frac{1}{2}
{\rm Tr}\Bigl[P\wedge* P+P'\wedge \ast {P'}\Bigr]
\nn\\
{\cal L}_{\rm WZ}&=&-{\cal K} {\rm tr}_B\Bigl[
{\Sigma}\wedge {S}+\Omega\wedge D+
{\Sigma'}\wedge {S'}+ \Omega'\wedge D'\Bigr]\nn\\
&=&-{\cal K} {\rm Tr}\Bigl[{\Sigma}\wedge {S}+\Omega\wedge D-
{\Sigma'}\wedge {S'}- \Omega'\wedge D'\Bigr]~~~.
\eea
From the MC equations in \bref{mcsuper0}
\bea
d(\sigma-s+\omega-{\rm d})
&=&-j\wedge j-{\rm d}\wedge{\rm d}-p\wedge{\rm d}-{\rm d}\wedge p\nn\\
&=&-d\Bigl(j+\frac{1}{2}(-\omega+{\rm d})\Bigr)\nn\\
&=&-d\Bigl(p+\frac{3}{2}{\rm d}+\frac{1}{2}\omega+s+\sigma\Bigr)~~~,
\eea
the equation of motion is obtained as
\bea
&&d\ast j_N=0~,~j_N=p-{\cal K}\ast
(p+\frac{3}{2}{\rm d}+\frac{1}{2}\omega+s+\sigma)
\nn\\
&&d\ast j_N'=0~,~j_N'=p'+
{\cal K}\ast(p'+\frac{3}{2}{\rm d}'+\frac{1}{2}\omega'+s'+\sigma')
\label{eomsust}
\eea
Similar structure was obtained for a superstring in the AdS space in
\cite{Bena:2003wd,Hatsuda:2005te}.
Since two sets of algebras are decoupled, 
equations of motion for double coordinates are also decoupled as;
\bea
&&{\cal K}=-1~\Rightarrow~\nn\\
&&{\renewcommand{\arraystretch}{1.6}
\left\{\begin{array}{ll}
\partial_+\left(p_-+\frac{1}{2}(s_-+{\sigma}_-+\frac{3}{2}{\rm d}_-+\frac{1}{2}\omega_-)
\right)
-\frac{1}{2}\partial_-(s_++\sigma_++\frac{3}{2}{\rm d}_++\frac{1}{2}\omega_+)=0
\\
\partial_-\left(p'_++\frac{1}{2}(s'_++{\sigma}'_++\frac{3}{2}{\rm d}'_++\frac{1}{2}\omega'_+)
\right)-
\frac{1}{2}\partial_+(s'_-+{\sigma}'_-+\frac{3}{2}{\rm d}'_-+\frac{1}{2}\omega'_-)
=0
\end{array}\right.}\nn\\
&&{\rm or}~~~\label{eqsust}\\
&&{\cal K}=1~\Rightarrow~\nn\\
&&
{\renewcommand{\arraystretch}{1.6}
\left\{\begin{array}{ll}
\partial_-\left(p_++\frac{1}{2}(s_++{\sigma}_++\frac{3}{2}{\rm d}_++\frac{1}{2}\omega_+)\right)-
\frac{1}{2}\partial_+(s_-+{\sigma}_-+\frac{3}{2}{\rm d}_-+\frac{1}{2}\omega_-)=0\\
\partial_+\left(p'_-+\frac{1}{2}(s'_-+{\sigma}'_-+\frac{3}{2}{\rm d}'_-+\frac{1}{2}\omega'_-)
\right)
-\frac{1}{2}\partial_-(s'_++\sigma'_++\frac{3}{2}{\rm d}'_++\frac{1}{2}\omega'_+)=0
\end{array}\right.}\nn
\nn
\eea
For ${\cal K}=1$  the same dimensional reduction conditions as the previous subsection, 
$S=S'=0$ and
$\sigma=\sigma'=0$, are imposed.
It is possible to choose chiral gauge ${\rm d}_-=\omega_-=0$
and ${\rm d}'_+=\omega'_+=0$
 as shown in \cite{Gates:1989hg,
Hatsuda:2008xa}.
Equations of motion are chiral in this gauge,  $\partial_- j_N=0$ and $\partial_+ j'_{N}=0$.

The equations of motion in flat space are the followings.
The following lightcone gauge  is imposed,
\bea
x^+=x^+_0+p^+(\sigma+\tau)~,~
x'{}^+=x'^+_0+p^+(\sigma-\tau)~,~
\gamma^+\theta=\gamma^+\theta'=0~~~,
\eea
with constant $p^+$.
Then the left/right conserved currents in \bref{eomsust} 
become chiral currents,  
\bea
\partial_+{\rm tr} (j_N P_{\perp} m)=0~\Rightarrow \partial_+(\partial_- x_{\perp m})=0~&,&~
\partial_+ {\rm tr}(j_N D_\mu)=0~\Rightarrow 
\partial_+(p^+\gamma^-\theta )_\mu=0\nn\\
\partial_-{\rm tr} (j'_N P_{\perp m'})=0~\Rightarrow \partial_-(\partial_- 
x_{\perp m'})=0~&,&~
\partial_-{\rm tr}(j'_N D_{\mu'})=0~\Rightarrow 
\partial_-(p^+\gamma^-\theta' )_{\mu'}=0\nn~~.
\eea
The dimensional reduction condition reduces $x$ and $x'$ 
into the usual one, for example $X=x+x'$,
then equations of motion reduce into the usual one
\bea
\partial_+\partial_- X^{\perp m}=0~,~\partial_+\theta^{\perp \mu }=0~,~
\partial_-\theta^{ \perp\mu'}=0
\eea
for the lightcone variables $X^{\perp m},~\theta^{\perp \mu},~\theta^{\perp \mu'}$.

\par\vskip 6mm

\subsection{Type II superstring with R-R extension }

In this subsection we include
the R-R central charges  $\Upsilon$ and $\digamma$. 
For the nondegenerate type II super-Poincar\'{e} algebra with R-R central  charge 
we use following notation of currents 
\bea
& J_{[0]}=\{S,~S'\}~,~J_{[\frac{1}{2}]}=\{D,~D'\}~,~
J_{[1]}=\{P,P',\Upsilon,\digamma\}~,~
J_{[\frac{3}{2}]}=\{\Omega,\Omega' \}~,~
J_{[2]}=\{\Sigma,~\Sigma'\}\nn~~~&\\
 &j_{[0]}=\{s,~s'\}~,~j_{[\frac{1}{2}]}=\{{\rm d},~{\rm d}'\}~,~
j_{[1]}= \{p,~p',~u, ~f\}~,~
j_{[\frac{3}{2}]}=\{\omega,~\omega'\}~,~
j_{[2]}=\{\sigma,~\sigma'\}\nn~~.&
\eea
The left-invariant currents and the MC equations are 
\bea
g^{-1}dg&=& {J}+J'+\Upsilon+\digamma~\nn\\
{J}&=&
 {S}+ {D}+ {P}+ {\Omega}+ {\Sigma}
~,~J'~=~ {S}'+ {D}'+ {P}'+ {\Omega}'+ {\Sigma}' 
 \nn\\
d {S}&=& -{S}\wedge  {S}\nn\\
d {D}&=& -{S}\wedge  {D}- {D}\wedge  {S}\nn\\
d {P}&=& -{S}\wedge  {P}- {P}\wedge  {S}- {D}\wedge  {D}\nn\\
d {\Omega}&=& -{S}\wedge  {\Omega}
-{\Omega}\wedge  {S}- {D}\wedge  {P}
- {P}\wedge  {D}-D'\wedge\digamma-\digamma\wedge D'
\nn\\
d {\Sigma}&=& -{S}\wedge  {\Sigma}
- {\Sigma}\wedge  {S}-{D}\wedge  {\Omega}
- {\Omega}\wedge  {D}-
 {P}\wedge  {P}-\frac{1}{2}\digamma\wedge \Upsilon-
 \frac{1}{2}\Upsilon\wedge\digamma
 ~~~\\
 d {\Upsilon}&=& -({S}+S')\wedge  {\Upsilon}-{\Upsilon}\wedge (S+S')
   -D\wedge D'-D'\wedge D\nn\\
 d {\digamma}&=& -({S}+S')\wedge  {\digamma}-{\digamma}\wedge (S+S')\nn\\
  0&=& \Sigma\wedge\Sigma=\Sigma\wedge \Omega+\Omega\wedge \Sigma=\Sigma\wedge P+P\wedge \Sigma=\Sigma\wedge \Upsilon+\Upsilon\wedge \Sigma
=\Sigma\wedge \digamma+\digamma\wedge \Sigma 
 \nn\\
 &=&\Sigma\wedge D+D\wedge \Sigma=\Omega\wedge \Omega=\Omega\wedge P+P\wedge \Omega
 =\Omega\wedge \Upsilon+\Upsilon\wedge \Omega
=\Omega\wedge\digamma+\digamma\wedge \Omega
 ~~~.\nn
\eea
The MC equations for primed left-invariant currents are  same as the above with replacing unprimed indices with primed indices.
Decomposed  right-invariant currents and the MC equations are
\bea
(dg)g^{-1}&=&{j}+j'+u+f~\nn\\
{j}&=&
 {s}+ {\rm d}+ {p}+ {\omega}+ {\sigma}
~,~j'~=~{s}'+ {\rm d}'+ {p}'+ {\omega}'+ {\sigma}'
 \nn\\
d {s}&=&j\wedge j-p\wedge p-{\rm d}\wedge {\rm d}
- {{\rm d}}\wedge (  {p}+ {\omega})
-({p}+ {\omega})\wedge {{\rm d}}+s\wedge(u+f)+(u+f)\wedge s
\nn\\
d {{\rm d}}&=&2 {{\rm d}}\wedge  {{\rm d}}
+ {{\rm d}}\wedge ( {p}+  {\omega}+{\rm d}'+f)+( {p}+  {\omega}+{\rm d}'+f)\wedge  {{\rm d}}
\nn\\
d {p}&=& -{{\rm d}}\wedge  {{\rm d}}+2 {p}\wedge  {p}
+ {{\rm d}}\wedge  {p}+ {p}\wedge  {{\rm d}}\nn\\
d {\omega}&=& {{\rm d}}\wedge(-  {p}+ {\omega})
+(  -{p}+ {\omega})\wedge {{\rm d}}-{\rm d}'\wedge f-
f\wedge {\rm d}'\nn\\
d {\sigma}&=& -{p}\wedge  {p}
- {{\rm d}}\wedge  {\omega}- {\omega}\wedge  {{\rm d}}
-\frac{1}{2}u\wedge f-\frac{1}{2}f\wedge u
~\label{mcsuper1}\\
d {u}&=& -{{\rm d}}\wedge  {{\rm d}}' -{{\rm d}}'\wedge  {{\rm d}}
+u\wedge f+f\wedge u\nn\\
d {f}&=& ({{\rm d}}+{\rm d}')\wedge f+f\wedge  ({{\rm d}}+{\rm d}')
+u\wedge f+f\wedge u\nn\\
0&=&\sigma\wedge\sigma=\sigma\wedge\omega+\omega\wedge\sigma=
\sigma\wedge p+p\wedge\sigma=\sigma\wedge u+u\wedge\sigma
=\sigma\wedge f+f\wedge\sigma
\nn\\
&=&\sigma\wedge {\rm d}+{\rm d}\wedge\sigma=\omega\wedge\omega
=\omega\wedge p+p \wedge\omega =\omega\wedge u+u \wedge\omega =
\omega\wedge f+f \wedge\omega \nn
~~~.
\eea
The MC equations for primed decomposed right-invariant currents are same as the above with replacing unprimed indices with primed indices.

An action for a superstring in manifestly T-duality formalism is given by
\bea
{\cal L}_0&=&-\frac{1}{2}
\hat{\rm tr}\Bigl[P\wedge* P+P'\wedge \ast {P'}\Bigr]~=~
-\frac{1}{2}
{\rm Tr}\Bigl[P\wedge* P+P'\wedge \ast {P'}\Bigr]
\nn\\
{\cal L}_{\rm WZ}&=&-{\cal K} {\rm tr}_B \Bigl[{\Sigma}\wedge {S}+\Omega\wedge D+
{\Sigma'}\wedge {S'}+ \Omega'\wedge D'+\digamma\wedge \Upsilon\Bigr]\nn\\
&=&-{\cal K} {\rm Tr}\Bigl[{\Sigma}\wedge {S}+\Omega\wedge D-
{\Sigma'}\wedge {S'}- \Omega'\wedge D'+\digamma\wedge \Upsilon\Bigr]~~~.
\eea
From the MC equations in \bref{mcsuper1}
\bea
&&{\rm tr}\Delta d(\sigma-s+\omega-{\rm d}+
\sigma'-s'+\omega'-{\rm d}'+f-u)\nn\\
&&~~=
-{\rm tr}\Delta d\left(j+j'+
\frac{1}{2}(-\omega+{\rm d}-\omega'+{\rm d}')+2u\right)\nn\\
&&~~=-{\rm tr}\Delta 
d\left(p+\frac{3}{2}{\rm d}+\frac{1}{2}\omega+s+\sigma
+p'+\frac{3}{2}{\rm d}'+\frac{1}{2}\omega'+s'+\sigma'
+2u
\right)~~~,
\eea
the equation of motion is obtained as
\bea
&&{\rm tr}\Delta~
d\ast\Bigl(p+p'-{\cal K}\ast
(p+\frac{3}{2}{\rm d}+\frac{1}{2}\omega+s+\sigma+
p'+\frac{3}{2}{\rm d}'+\frac{1}{2}\omega'+s'+\sigma'
+2 u)\Bigr)=0\nn\\
\eea
After imposing the constraint $\Upsilon=0$  leading to $u=0$
in addition to $S=S'=0$ and $\sigma=\sigma'=0$ constraints,
equations of motion reduces to the one obtained in \bref{eqsust}.
\par\vskip 6mm

\section{Conclusions and discussions}

We present a superspace formulation of type II superstring background 
with manifest T-duality  symmetry including R-R gauge fields.
The nondegenerate super-Poincar\'{e} affine algebra with the central extension 
is given  in \bref{SDPomesig} and \bref{RRalg} to define the superspace.
The $\kappa$-symmetric Virasoro constraints are obtained.
It determines torsion constraints confirming the consistency with the Bianchi identities. 
Obtained torsion constraints are much simpler than the previous case \cite{Hatsuda:2014qqa}:
All torsions with dimension 1 and less are trivial.
The vielbein superfield $E_{\underline{AM}}$ is  identified with the supergravity multiplet.
The R-R gauge fields are $C_{{\rm RR}\underline{a}}{}^{\alpha\alpha'}=E_{\underline{a}\digamma}$ and R-R gauge field strengths are
$F_{\rm RR}{}^{\alpha\alpha'\underline{bc}}=E_{\digamma\Sigma}$ as well as
$F_{\rm RR}{}^{\alpha\beta'}=E_{\Omega\Omega'}$,
while NS-NS gauge field strengths are  
$F_{\rm NS}{}_{\underline{a}}{}^{\underline{bc}}=E_{\underline{a\Sigma}}$ 
as well as
$F_{\rm NS}{}^{\alpha\beta}=E_{\Omega\Omega}$ and 
$F_{\rm NS}{}^{\alpha'\beta'}=E_{\Omega'\Omega'}$.

A Green-Schwarz superstring action is given where 
the Wess-Zumino term is written by bilinears of left-invariant currents locally.
Therefore the Noether current is obtained easily.
Left and right moving currents are separated by the auxiliary Lorentz coordinates. 
Auxiliary degrees of freedom of the space are removed by dimensional reduction conditions
in addition to the section condition.

It is also possible to write down 
an action in the Hamiltonian formalism 
for a superstring in manifestly T-duality formalism 
with manifest $\kappa$ symmetry is given as
\bea
{\cal L}&=&\partial_\tau Z^{\underline{M}}\partial_{\underline{M}}
-\chi_\tau{\cal H}_\tau-\chi_\sigma{\cal H}_\sigma
-\chi_\alpha{\cal B}^\alpha-\chi_{\alpha'}{\cal B}^{\alpha'}\nn\\&&
-\chi_1{\cal C}-\chi_2{\cal D}-\chi_3{\cal C}'-\chi_4{\cal D}
+\chi_SS+\chi_{S'}S'+\chi_\Upsilon\Upsilon
~~~
\eea
$\chi$'s are all multipliers of the constraints.
The section condition and the strong condition  
in the doubled coordinate space are
\bea
\partial^{\underline{M}}\partial_{\underline{M}}~\Psi(Z^{\underline{L}})=
\partial^{\underline{M}}\Phi(Z^{\underline{L}})~\partial_{\underline{M}}\Psi(Z^{\underline{L}})=
0\label{section}
\eea
for physical states $\Phi(Z^{\underline{L}})
$ and $\Psi(Z^{\underline{L}})$.

There are many interesting questions which should be clarified such as
structure of supersymmetric AdS space,  branes and exotic branes in the superspace
and the appearance of the minimum length.

\par\vskip 6mm

\section*{Acknowledgements}
W.S. thanks  to
 Martin Pol\'{a}{\v{c}}ek and 
 M.H. thanks to Jeong-Hyuck Park for valuable discussions. 
 M.H. and K.K. also thanks Yukawa Institute for Theoretical Physics at Kyoto University
 for useful talks and discussions at the YITP workshop on
 "Exotic Structures of Spacetime" in March, 2014.
M.H. would like to thank the Simons Center for Geometry and Physics for
hospitality during ``the 
2014 Summer Simons workshop in Mathematics and Physics"
where this work has been developed. 
 The work of M.H. is supported  by Grant-in-Aid for Scientific Research (C)
  No. 24540284 from The Ministry of Education, Culture, Sports, Science and Technology of Japan,
and the work of W.S. is
 supported in part by National Science Foundation Grant No. PHY-1316617.


\end{document}